\documentclass[pdflatex,sn-mathphys-num]{sn-jnl}% Math and Physical Sciences Numbered Reference Style
\usepackage{float}
\usepackage{comment}
\usepackage{graphicx}
\usepackage{subcaption}
\usepackage{multirow}%
\usepackage{amsmath,amssymb,amsfonts}%
\usepackage{amsthm}%
\usepackage{mathrsfs}%
\usepackage[title]{appendix}%
\usepackage{xcolor}%
\usepackage{textcomp}%
\usepackage{manyfoot}%
\usepackage{booktabs}%
\usepackage{algorithm}%
\usepackage{algorithmicx}%
\usepackage{algpseudocode}%
\usepackage{listings}%
\usepackage[switch]{lineno}
\theoremstyle{thmstyleone}%
\theoremstyle{thmstyletwo}%

\theoremstyle{thmstylethree}%

\begin{document}
%TC:ignore
\title[Settlement percolation reveals shrinking critical distances across the urban hierarchy]{Settlement percolation reveals shrinking critical distances across the urban hierarchy}


\author*[1]{\fnm{Somwrita} \sur{Sarkar}}\email{somwrita.sarkar@sydney.edu.au}

\author[1]{\fnm{Chirag} \sur{Deb}}

\author[1]{\fnm{Carolina} \sur{Pantoja Mardones}}

\author[3]{\fnm{Anmol} \sur{Gaur}}

\author[3]{\fnm{Arunim} \sur{Anand}}

\author[1]{\fnm{Rashi} \sur{Shrivastava}}

\author[3]{\fnm{Anusha} \sur{Roy}}

\author[3]{\fnm{Arnab} \sur{Jana}}

\author[3]{\fnm{Eswar} \sur{Rajasekaran}}

\author[2]{\fnm{David} \sur{Levinson}}


\affil*[1]{\orgdiv{School of Architecture, Design and Planning}, \orgname{The University of Sydney}, \orgaddress{\street{148, City Road}, \city{Darlington}, \postcode{2008}, \state{NSW}, \country{Australia}}}

\affil[2]{\orgdiv{School of Civil Engineering}, \orgname{The University of Sydney}, \orgaddress{\street{225 Shepherd Street}, \city{Darlington}, \postcode{2008}, \state{NSW}, \country{Australia}}}

\affil[3]{\orgdiv{Department of Civil Engineering}, \orgname{Indian Institute of Technology, Bombay}, \orgaddress{\street{Powai}, \city{Mumbai}, \postcode{400076}, \state{Maharashtra}, \country{India}}}


\abstract{Measures of urban expansion quantify land consumption but obscure how separated settlements coalesce. We examine India's 50 largest urban agglomerations (1975–2025) using percolation-like analysis. Critical distances marking the largest settlement-merger events contracted in 47 cities, the median falling 46\% from 650 to 350m. This longitudinal shrinking of settlement connectivity length scales reveals a morphological mechanism shaping land-consumptive settlement growth through distinct forms of inter-cluster coalescence (compact, corridor, satellite, regional). Population size shows no detectable linear relationship with contraction magnitude, although intermediate-sized agglomerations display the widest observed spread. Confirming established land-consumption patterns, built-up surfaces and urban footprints grow faster than population. Yet their relative growth flips with urban delineation: before the largest merger, footprints outgrow built-up surface (386\% versus 361\%); immediately afterward, built-up surface outgrows footprints (315\% versus 287\%). Recognizing merger dynamics can guide infrastructure investment, prioritize sustainable infill, reduce land consumption, and protect agricultural and ecologically sensitive landscapes.}




\maketitle
%\linenumbers
%TC:endignore
\section{Introduction}\label{sec1}

The sustainability of urbanization depends not only on how much land settlements occupy, but also on how that land is spatially organized as settlements expand and coalesce. The United Nations (UN) World Urbanization Prospects 2025 estimates that nearly 80\% of the global population lives in cities, towns, and semi-urban areas~\cite{un2025}. Monitoring the growth of these settlements is central to Sustainable Development Goal (SDG) Indicator 11.3.1, which measures the ratio of the land consumption rate to the population growth rate. Yet conventional measures of urban expansion reveal how much land is consumed without necessarily explaining how separated settlements become connected over time. Moreover, measuring land consumption requires answering a fundamental question: what spatial object constitutes the ``city''?

This question reflects the modifiable areal unit problem (MAUP): measured spatial relationships depend on the scale and configuration of the units used for analysis~\citep{openshaw1977geographical}. As emphasized in the UN-Habitat World Cities Report 2022, inconsistent urban definitions undermine comparisons across places and over time~\cite{un2022}. The problem is particularly consequential for expanding metropolitan regions, where formerly separate towns become connected, development extends along infrastructure corridors, and dispersed settlements form larger regional systems. Administrative jurisdictions, census-defined cities, and remotely-sensed urban footprints may therefore describe different geographies. Even within the same settlement data, changing the spatial connectivity criterion can alter which settlements belong to a single urban entity, with direct consequences for measured land consumption and population growth. Urban delineation is thus not merely a technical prerequisite for measuring expansion; it shapes the phenomenon being measured.

India provides an important setting for examining these processes. Its last completed population census was conducted in 2011, leaving a substantial gap in official observations of subsequent urban transformation. Moreover, census-based urban classifications do not necessarily capture the spatial and functional integration of settlements. Under India's census-town criterion, at least 75\% of \emph{male main workers} must be employed in non-agricultural activities, excluding women's employment from this test and overlooking marginal workers, including many seasonal and temporary migrants. Such historically gendered criteria are particularly consequential in expanding urban peripheries, where settlements may remain administratively rural while becoming part of larger urban systems. Defining the urban therefore affects both sustainability measurement and the recognition of populations and places within urban policy.

We examine India's 50 largest urban agglomerations, selected using 2025 population projections from the UN World Urbanization Prospects 2018 Revision~\cite{un2018,un2018_2}. Together, they account for approximately 74\% of the population of Indian urban agglomerations represented in that dataset. Although identifiable in population statistics, these agglomerations lack a consistent set of official spatial boundaries suitable for longitudinal comparison~\cite{populationdata2011}. Their evolving spatial extent and connectivity must therefore be reconstructed empirically.

Settlement coalescence need not follow a uniform process of outward expansion. Development is spatially selective: locations with better access to valued opportunities may attract households, firms, and investment. Gravity-based accessibility represents this attraction by weighting surrounding opportunities according to the impedance of reaching them~\citep{levinson2020towards}, while the co-evolution of transport networks and land use can reinforce spatially uneven development~\citep{levinson2026general}. Depending on transport connectivity, land availability, development costs, regulation, and environmental constraints, these processes may favor compact consolidation, growth along corridors, satellite absorption, or the emergence of regionally connected settlement systems. Improved access may also be expressed through redevelopment, greater building intensity, or higher land prices rather than additional occupied land. These are plausible mechanisms underlying different settlement morphologies; our analysis examines their spatial expression without estimating their causal determinants.

\begin{figure}[t]
\centering
\includegraphics[width=\textwidth]{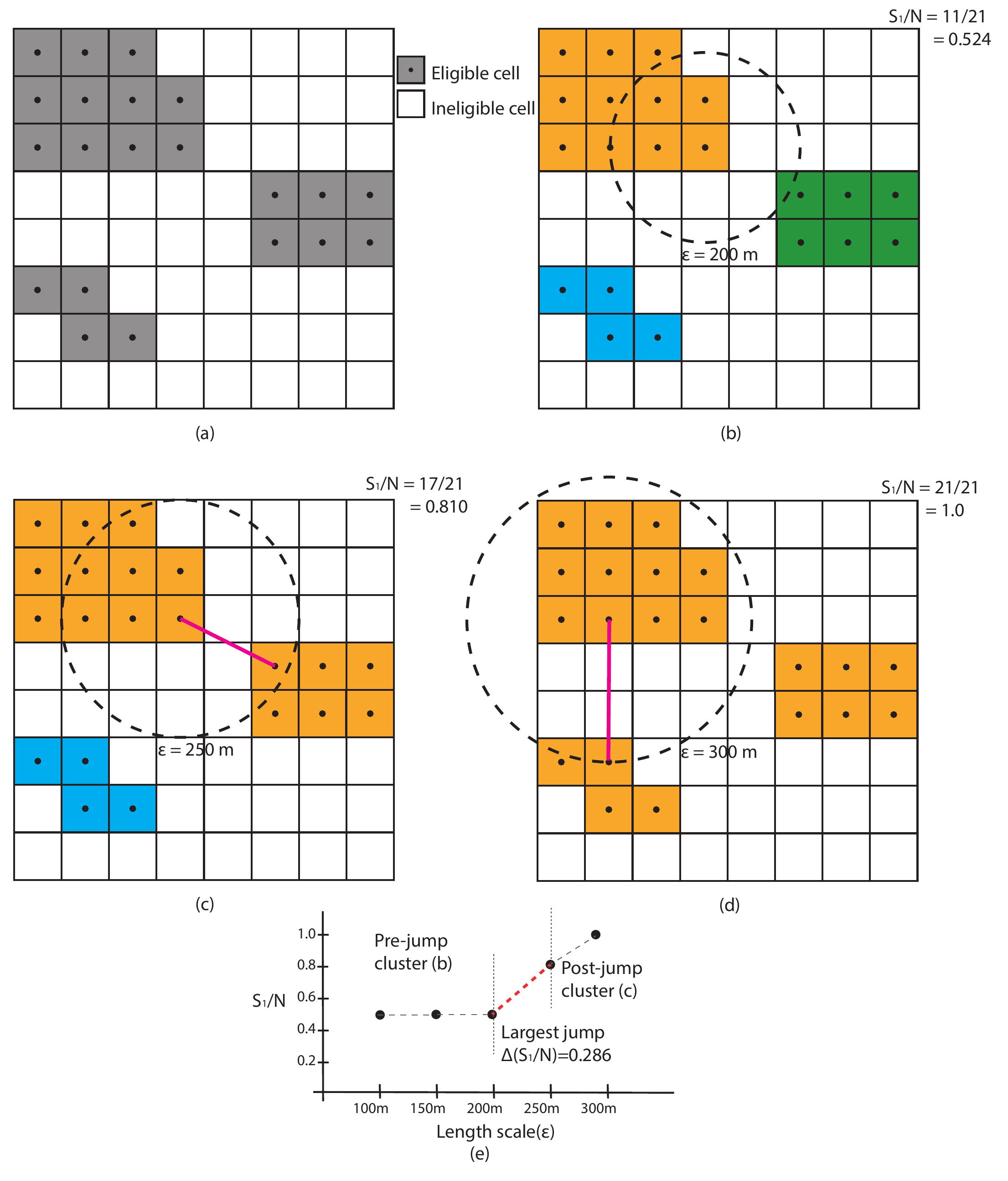}
\caption{\textbf{Schematic illustration of spatial percolation and the City Clustering Algorithm.}
(a) A synthetic settlement configuration comprising 21 eligible cells on a regular $100\,\mathrm{m}\times100\,\mathrm{m}$ raster grid. Eligible cells have nonzero population and built-up surface area, and their centers constitute the vertices of a spatial connectivity graph.
(b) At a neighborhood length of $\epsilon=200$ m, the eligible cells form three disconnected components containing 11, 6, and 4 cells, respectively. The largest-cluster share is $S_1/N=11/21=0.524$.
(c) Increasing $\epsilon$ to 250 m permits connections between previously disconnected components, producing a largest cluster of 17 cells, with $S_1/N=17/21=0.810$.
(d) At $\epsilon=300$ m, the remaining component becomes connected, producing a single cluster containing all 21 eligible cells, with $S_1/N=1$.
(e) Evolution of the largest-cluster share, $S_1/N$, with increasing neighborhood length. The largest observed increase occurs between 200 m and 250 m, with $\Delta(S_1/N)=6/21=0.286$, identifying the characteristic merger interval. The corresponding configurations are designated the pre-jump and post-jump clusters. Dashed circles illustrate the neighborhood radius of a representative eligible cell, while highlighted links indicate newly permitted connections between previously disconnected components. All connected components are identified using unrestricted DBSCAN ($\mathrm{minPts}=1$).}
\label{fig:percolation_demo}
\end{figure}

To investigate how settlement connectivity evolves, we extend the City Clustering Algorithm (CCA)~\cite{rozenfeld2008laws,rozenfeld2011area} through a percolation-like analysis of settlement configurations. Eligible settlement cells are represented as vertices of a geometric graph, with connections between cell centers separated by no more than a neighborhood distance, $\epsilon$. As $\epsilon$ increases, previously disconnected components merge into progressively larger clusters. We track the proportion of eligible cells belonging to the largest component, $S_1/N$, and identify the largest observed increment as the characteristic settlement-merger interval. The corresponding connection distance provides an operational measure of the spatial scale at which a major reorganization of settlement connectivity occurs, rather than an assumed universal critical threshold. Figure~\ref{fig:percolation_demo} illustrates this transition.

Using the Global Human Settlement Layer population and built-up surface products (GHS-POP and GHS-BUILT-S)~\cite{schiavina_etal_2023_ghs_pop,pesaresi_politis_2023_ghs_built_s}, we reconstruct settlement connectivity at 100 m resolution across eleven observations from 1975 to 2025. Within study regions encompassing the wider district contexts of each agglomeration, we apply unrestricted DBSCAN ($\mathrm{minPts}=1$) to cells with positive population and built-up surface, varying $\epsilon$ from 100 to 2000 m in 50 m increments. This approach permits characteristic connection distances to emerge separately for each city and year without imposing a uniform population-density threshold. It also allows us to examine how the same settlement configuration produces different urban delineations immediately below and above its largest merger event.

We address three linked questions:
\begin{itemize}
    \item First, how have characteristic settlement connectivity lengths changed over five decades, and what distinct morphological pathways of settlement coalescence do these changes reveal?
    \item Second, does city population size systematically explain the magnitude or variability of contraction in these characteristic connection lengths?
    \item Third, how do connectivity-dependent urban delineations alter estimates of land consumption relative to population growth under SDG 11.3.1, particularly when distinguishing built-up surface from occupied urban footprint?
\end{itemize}

Our principal contribution is to identify a widespread contraction in the characteristic spatial scale of settlement coalescence, revealing a morphological dimension of urban growth that aggregate land-consumption measures do not capture. Across 50 agglomerations, the characteristic post-merger connection distance declined in 47 cities between 1975 and 2025, with the median falling from 650 to 350 m. The magnitude of this contraction shows virtually no linear dependence on the population of the largest cluster, although its observed variability differs across the urban hierarchy. Distinct forms of settlement coalescence reveal how land-consumptive growth can reorganize spatial connectivity through multiple morphological pathways. At the same time, measuring land consumption immediately before and after the largest merger can yield different diagnoses of the relative growth of urban footprints and built-up surfaces. Together, these findings establish shrinking settlement connectivity lengths as an important morphological feature of urban transformation, while demonstrating that the delineation of urban settlements fundamentally affects how their sustainability is assessed.

\clearpage

\section{Results}

\subsection{Percolation provides a consistent approach to delineating cities}
\label{results1}

\begin{figure*}[t]
\centering

% (a) Delhi 1975, pre-jump
\begin{subfigure}[t]{0.48\textwidth}
    \centering
    \includegraphics[
        width=\linewidth,
        trim={0 0.55in 0 0.55in},
        clip
    ]{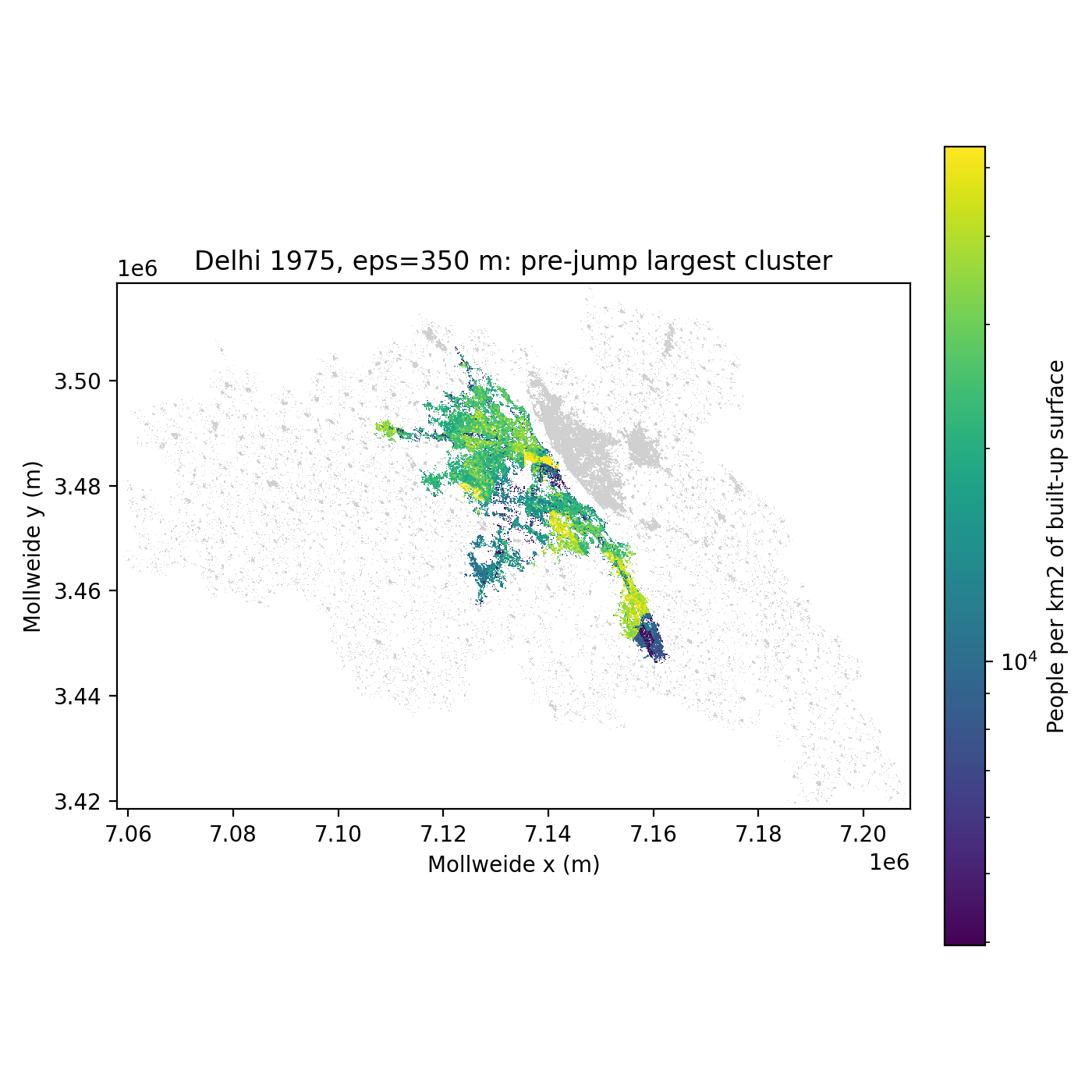}
    \caption{Delhi 1975, pre-jump ($\epsilon=350$~m)}
\end{subfigure}
\hfill
% (b) Delhi 1975, post-jump
\begin{subfigure}[t]{0.48\textwidth}
    \centering
    \includegraphics[
        width=\linewidth,
        trim={0 0.55in 0 0.55in},
        clip
    ]{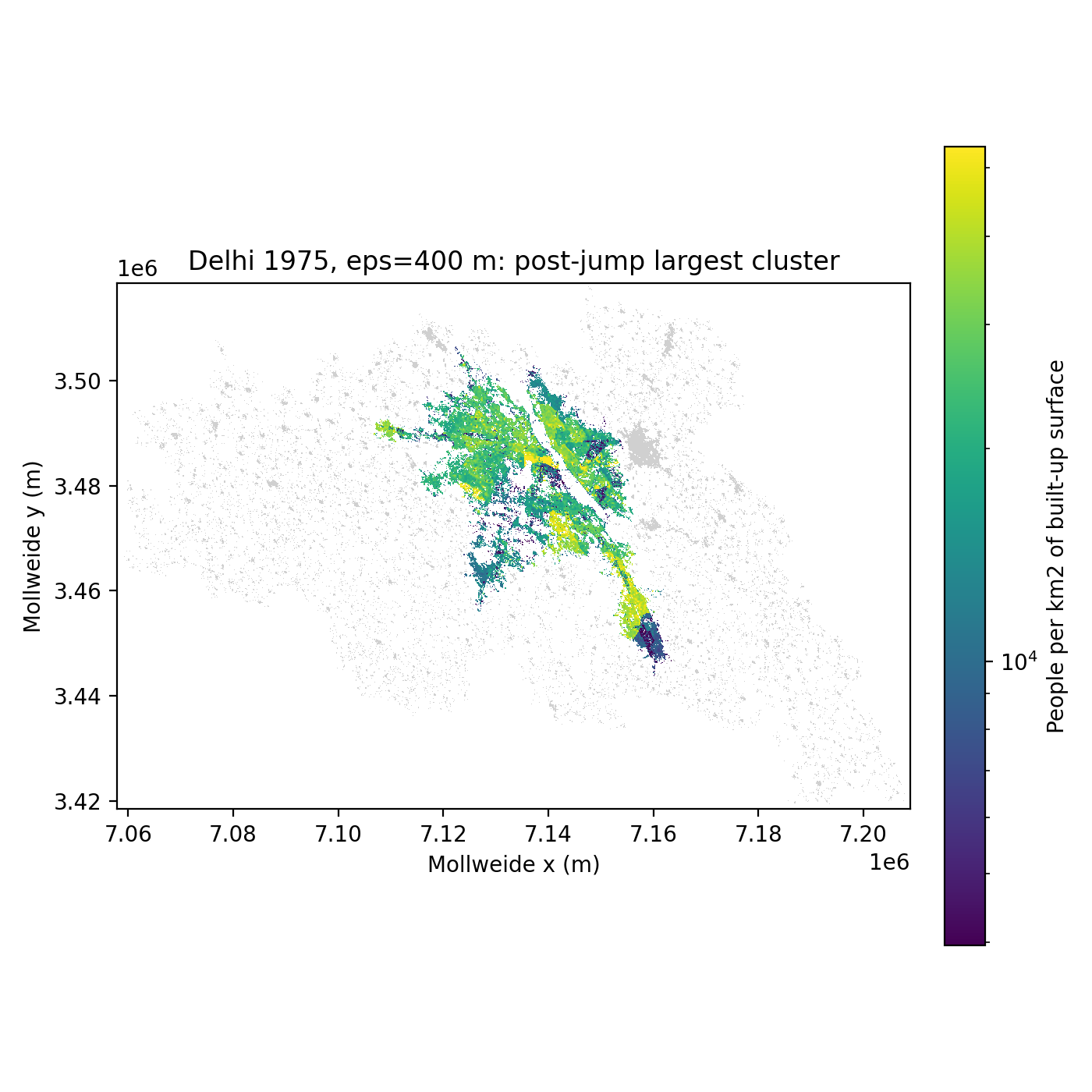}
    \caption{Delhi 1975, post-jump ($\epsilon=400$~m)}
\end{subfigure}

\vspace{0.3em}

% (c) Top ten clusters, Delhi 1975
\begin{subfigure}[t]{0.48\textwidth}
    \centering
    \includegraphics[
        width=\linewidth,
        trim={0 0.55in 0 0.55in},
        clip
    ]{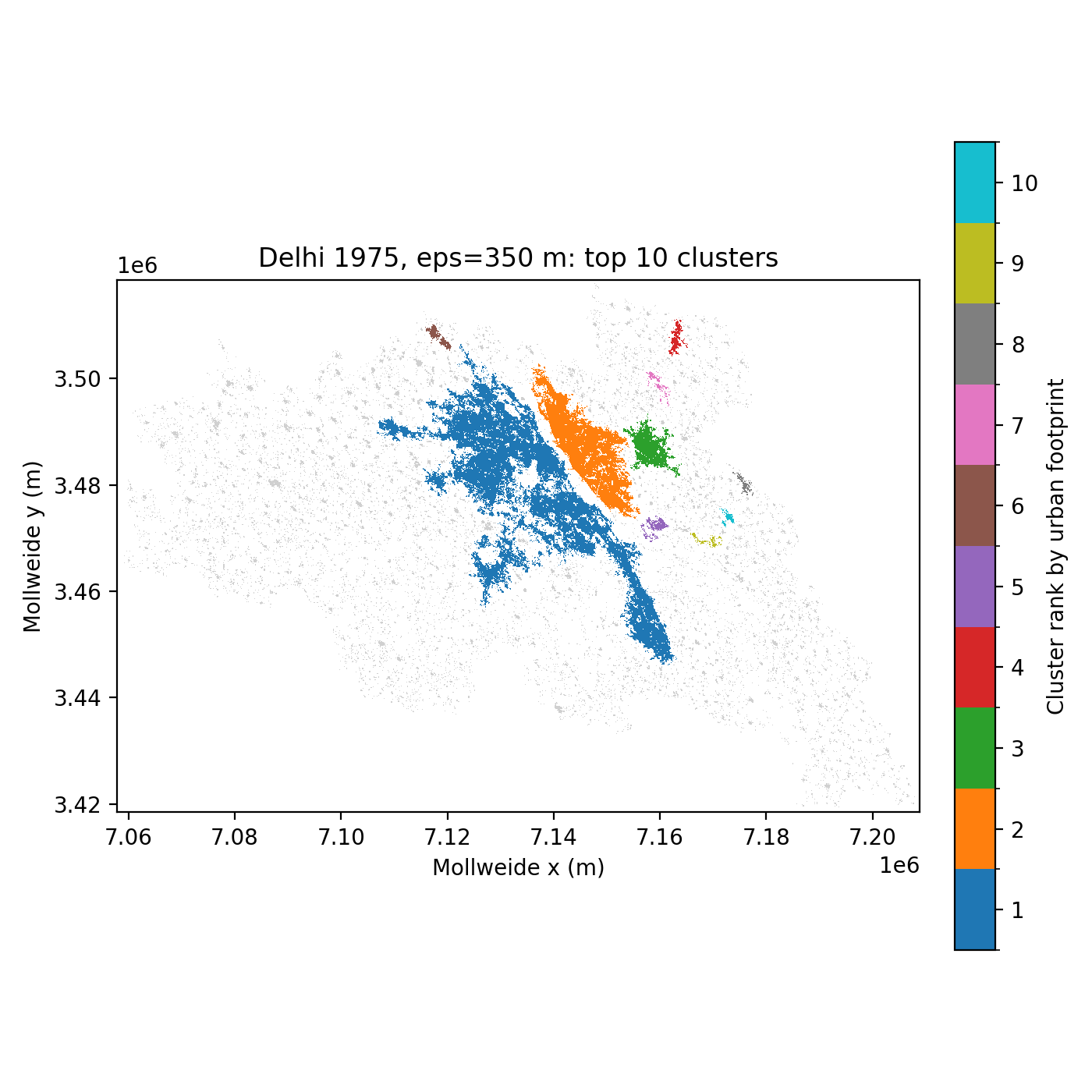}
    \caption{Ten largest clusters, Delhi 1975}
\end{subfigure}
\hfill
% (d) Connectivity curves, Delhi 1975--2025
\begin{subfigure}[t]{0.5\textwidth}
    \centering
    \includegraphics[
        width=\linewidth
    ]{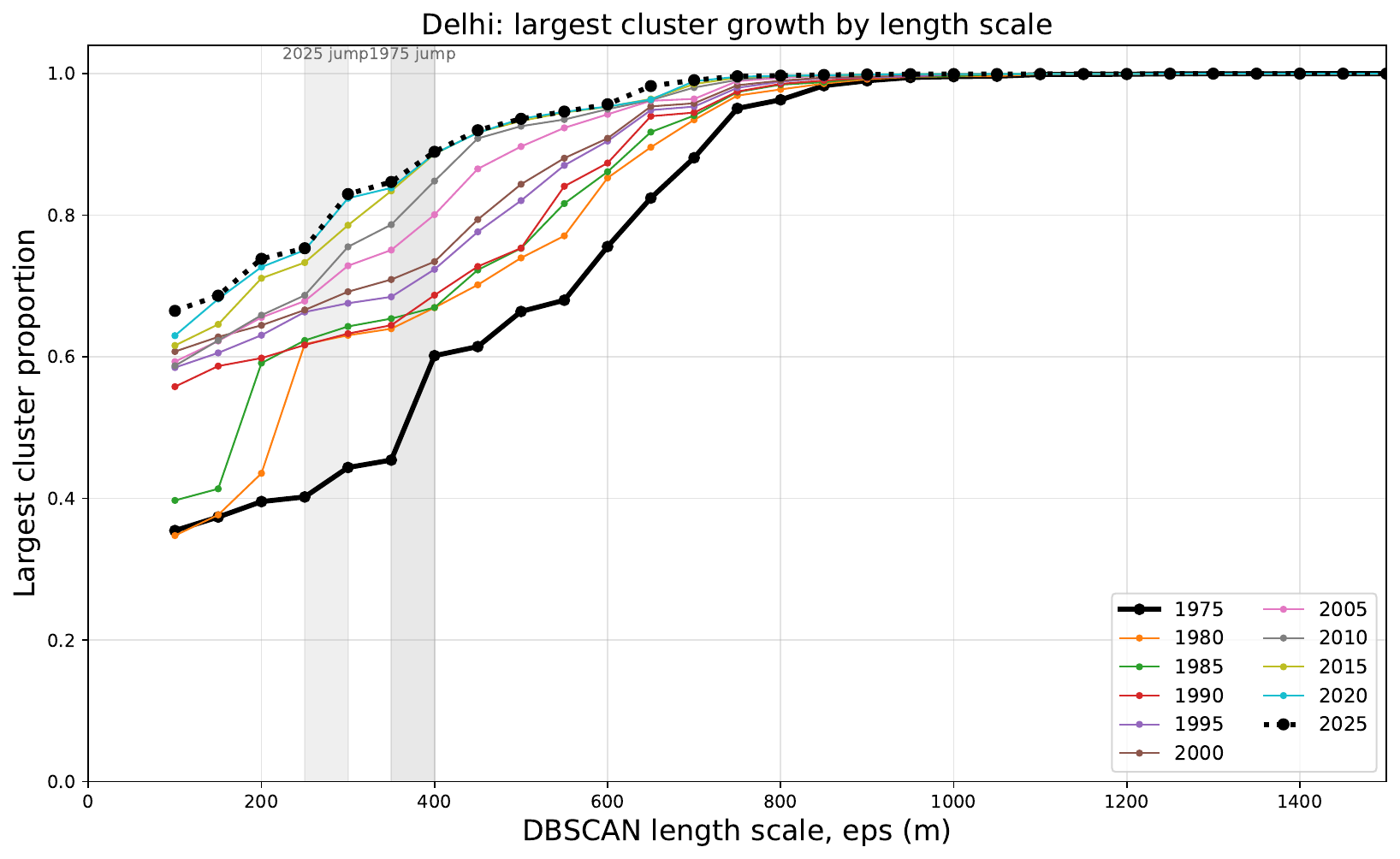}
    \caption{Largest-cluster share across connection lengths}
\end{subfigure}

\vspace{0.3em}

% (e) Pre/post-jump density distributions
\begin{subfigure}[t]{0.48\textwidth}
    \centering
    \includegraphics[
        width=\linewidth
    ]{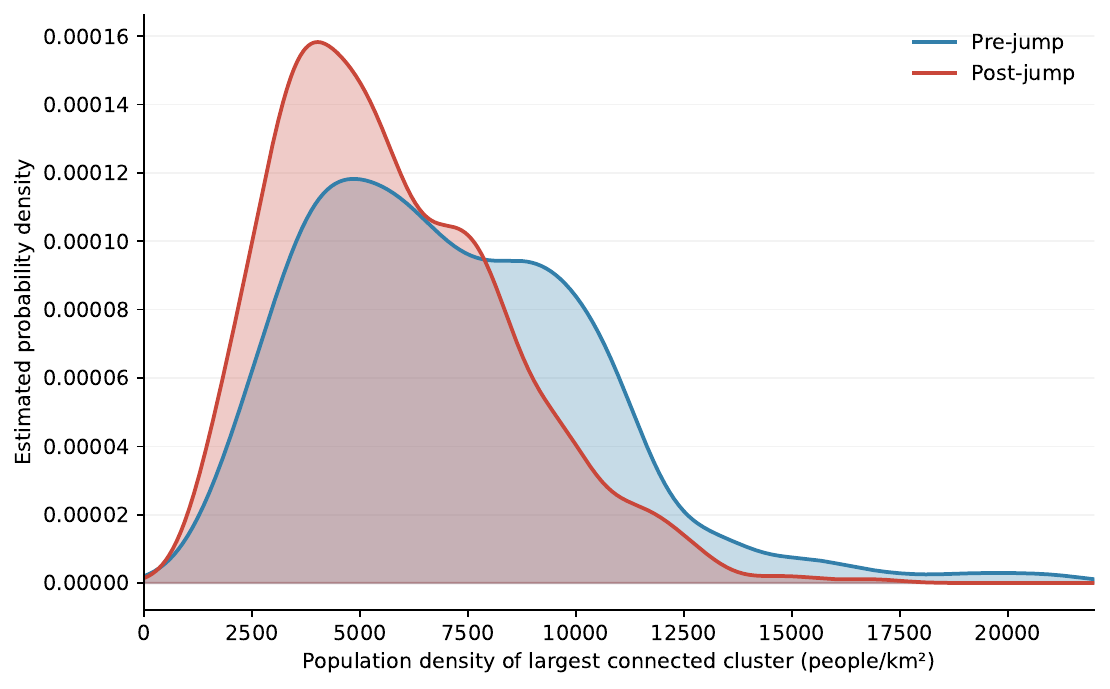}
    \caption{Population densities before and after major mergers}
\end{subfigure}
\hfill
% (f) 2025 grid-cell population density
\begin{subfigure}[t]{0.48\textwidth}
    \centering
    \includegraphics[
        width=\linewidth
    ]{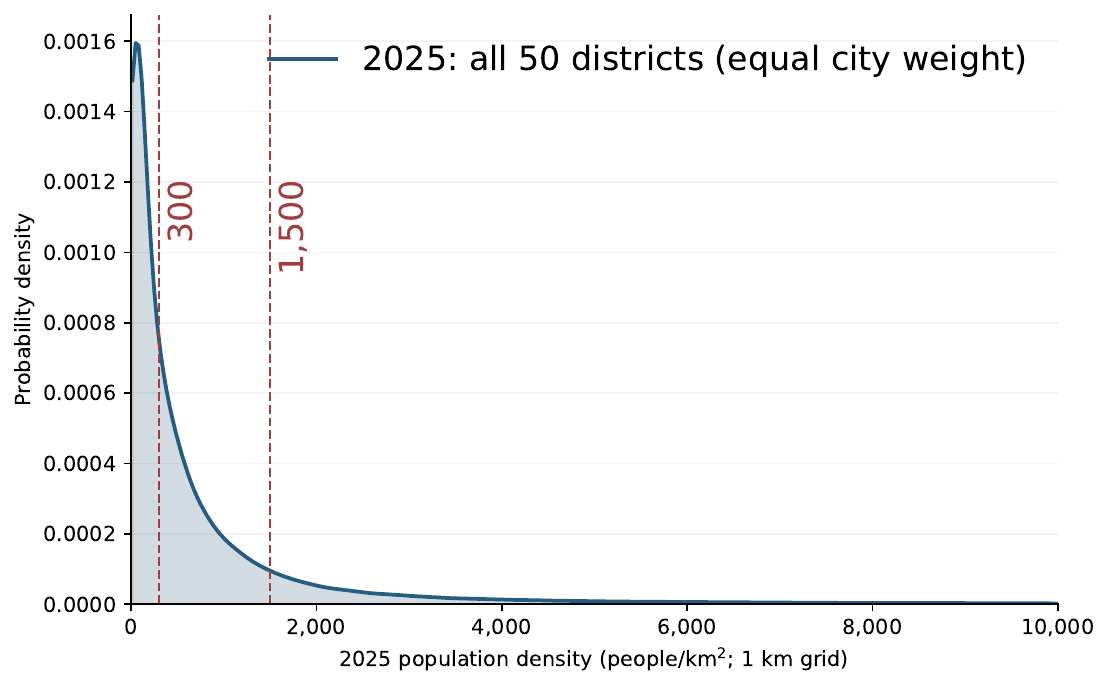}
    \caption{Grid-cell population densities, 2025}
\end{subfigure}

\caption{
\textbf{Percolation reveals major transitions in
urban connectivity without prescribing universal
population-density thresholds.}
(a,b) Largest connected cluster in Delhi in
1975, immediately before ($\epsilon=350$~m)
and after ($\epsilon=400$~m) the largest
observed merger.
(c) Ten largest settlement clusters in
Delhi in 1975 at $\epsilon=350$~m,
distinguished by cluster rank.
(d) Largest-cluster share, $S_1/N$, as a
function of DBSCAN neighbourhood length
$\epsilon$, for eleven observations
between 1975 and 2025. Shaded intervals
identify the largest observed 50~m
increases in $S_1/N$ for 1975 and 2025.
(e) Smoothed probability density distributions
of the mean population density of the largest
connected cluster immediately before and
after the largest merger, pooled across
50 cities and eleven observation years
($n=550$ for each distribution). Density
is measured per unit of connected urban
footprint.
(f) Equally weighted probability density
distribution of population density across
populated 1~km$^2$ grid cells within the
complete city-region administrative districts of all
50 cities in 2025. Dashed vertical lines
indicate the DEGURBA reference thresholds
of 300 and 1,500 people/km$^2$.
All clustering results use the unrestricted
specification ($\mathrm{minPts}=1$).
}
\label{fig:percolation_delhi}

\end{figure*}

We first examine how connected urban settlement systems emerge as the neighborhood length $\epsilon$ increases. Figure~\ref{fig:percolation_delhi} illustrates this process for Delhi. In 1975, the largest observed merger occurs between $\epsilon=350$ and 400 m (Figure~\ref{fig:percolation_delhi}(a--c)), increasing the largest-cluster share, $S_1/N$, from 45.4\% to 60.2\%, a gain of 14.8 percentage points. The ten largest clusters immediately before this merger (Figure~\ref{fig:percolation_delhi}(c)) illustrate how initially distinct settlement components subsequently coalesce. Applying the same procedure to Delhi in 2025 identifies a merger between 250 and 300 m, increasing $S_1/N$ from 75.3\% to 83.0\%, a gain of 7.7 percentage points. Such transitions arise either through the merger of substantial settlement components or the absorption of smaller components into an already dominant cluster.

Major mergers also alter the population density of the connected footprint. Across all 550 city-year observations, the pre-jump density distribution exhibits greater dispersion, whereas the post-jump distribution is more concentrated and shifted towards lower densities (Figure~\ref{fig:percolation_delhi}(e)). Median population density declines from 6,604 to 5,281 people/km$^2$, while the interquartile range contracts from 4,804 to 3,758 people/km$^2$. Population density decreases in 528 observations (96\%), consistent with the incorporation of relatively less densely populated peripheral settlements into larger connected systems. The emergence of more extensive connected urban footprints therefore need not reflect increasing population concentration, but may instead result from the integration of surrounding regional settlement fabrics.

To investigate the relevance of fixed density thresholds,
Figure~\ref{fig:percolation_delhi}(f) examines the distribution
of population density across populated 1 km$^2$ grid cells
within the complete district-based study regions, with each
of the 50 city study regions contributing equally to the
resulting probability density distribution, irrespective of
its geographical extent or number of populated grid cells. The distribution is strongly concentrated at low densities, declining progressively towards higher values. Neither the 300 people/km$^2$ threshold used by the Degree of Urbanization (DEGURBA) methodology to identify urban clusters (minimum population 5,000), nor the 1,500 people/km$^2$ threshold used to identify urban centers (minimum population 50,000), corresponds to an evident discontinuity or separate mode in the observed distribution~\cite{undesa2025wupmethodology}.

% Just noting that the following is incorrect - it is DEGURBA that specifies they use national thresholds, not statistical analysis. We are simply saying that there is no basis to support the 300 or 1500 because it comes from DEGURBA's own admission of using administrative data. 
%Comment BEGIN UNSDG-REVIEW-07
% Density thresholds: distinguish a visual comparison from a statistical test (Section 2.1).
%
% A smoothed distribution without a visible break does not establish the absence of a statistically
% significant boundary. Suggested replacement for the selected sentence:
%
% 'Neither prescribed density threshold corresponds to a visible discontinuity in the density distribution
% examined here.'
%
% Retain the existing DEGURBA methodology citation, \cite{undesa2025wupmethodology}, with the preceding
% definitions of the thresholds. No new citation can substitute for an unperformed significance test. The
% visual comparison also does not, by itself, invalidate thresholds designed for standardized international
% classification. If a formal discontinuity or multimodality test was performed, describe its method,
% uncertainty, and results before using significance language.
%Comment END UNSDG-REVIEW-07

These results distinguish two approaches to urban delineation. DEGURBA employs standardized population-density and population-size thresholds, informed partly by existing national urban definitions and designed to facilitate international comparison. However, these thresholds need not correspond to the underlying structure of observed settlement-density distributions and may exacerbate MAUP-related sensitivity in estimates of urban extent and land consumption. Percolation-like approaches, by contrast, systematically identify the connection lengths at which spatially separated settlements merge into larger systems. The resulting merger intervals provide a reproducible, empirically derived basis for distinguishing dominant connected urban components from more extensive regional settlement fabrics, without prescribing a universal population-density threshold at which urbanity begins. They also establish a consistent basis for examining how these characteristic connection lengths change across cities and over time.

\clearpage

\subsection{Percolation reveals a systematic reduction in characteristic connection lengths}

The characteristic connection length at which the largest merger occurs decreases systematically across India's 50 largest urban agglomerations between 1975 and 2025. Figure~\ref{fig:shrinking_length}(a) shows that the median post-jump length declines from 650 m to 350 m, a reduction of approximately 46\%, while the interquartile range contracts from 512.5--750 m to 250--450 m. This reduction is widespread: 47 cities exhibit shorter connection lengths, three remain unchanged, and none records an increase (Figure~\ref{fig:shrinking_length}(b)). The median reduction across the 50 individual cities is 250 m. 

At a fixed connection distance of 350 m, the median proportion of settlement cells belonging to the largest cluster increased from 20.6\% in 1975 to 64.7\% in 2025, with increases observed in 49 of 50 cities.

This temporal shift is illustrated in Delhi's largest-cluster trajectories (Figure~\ref{fig:percolation_delhi}(d)). At a fixed connection length of $\epsilon=100$ m, the largest-cluster share increases from 35.5\% in 1975 to 66.5\% in 2025, while at $\epsilon=1000$ m, both periods approach complete connectivity (99.6\% and 99.9\%, respectively). Thus, the dominant settlement component becomes established at shorter connection distances over time.

To test whether this finding depends on the largest-jump criterion, we examine alternative connectivity thresholds, defined as the smallest sampled length $\epsilon_q$ at which $S_1/N$ reaches a specified proportion $q$. Table~\ref{tab:connectivity_robustness} shows that the median $\epsilon_{50}$ declines from 650 m to 275 m, and $\epsilon_{75}$ from 750 m to 450 m, with reductions in all 50 cities. The median $\epsilon_{90}$ also declines, from 825 m to 600 m, with reductions in 47 cities. At a fixed $\epsilon=400$ m, the median largest-cluster share increases from 22.1\% to 73.6\%, with increases in 49 cities. The shortening of connection lengths is therefore robust to alternative measures of spatial connectivity.

The magnitude of the largest observed merger also declines, from a median of 25.7 percentage points in 1975 to 18.2 percentage points in 2025, with reductions in 40 cities (Table~\ref{tab:connectivity_robustness}). This suggests that connectivity transitions have become more diffuse in many urban systems as larger settlement components become established at shorter distances, consistent with the progressive reduction of spatial gaps between previously disconnected settlement components. However, increasing connectivity does not necessarily imply increasing population density or spatial compactness. Rather, it may reflect both the consolidation of existing settlement fabrics and the progressive integration of surrounding regional landscapes.

\begin{table}[t]
\centering
\caption{\textbf{Robustness of declining spatial connection lengths
across India's 50 largest urban agglomerations, 1975--2025.}
The largest-jump scale is the post-jump connection length identified
by the maximum increment in $S_1/N$. The alternative thresholds
$\epsilon_{50}$, $\epsilon_{75}$, and $\epsilon_{90}$ denote the
smallest sampled connection lengths at which the largest component
contains at least 50\%, 75\%, and 90\% of eligible settlement
cells, respectively. Values are cross-sectional medians across
50 cities in each year. City counts report the direction of
paired changes between 1975 and 2025. All results use unrestricted
DBSCAN ($\mathrm{minPts}=1$).}
\label{tab:connectivity_robustness}
\small
\begin{tabular}{lrrl}
\toprule
Measure & 1975 & 2025 & Cities with change \\
\midrule
Largest-jump scale (m) & 650 & 350 & 47 decreased \\
$\epsilon_{50}$ (m) & 650 & 275 & 50 decreased \\
$\epsilon_{75}$ (m) & 750 & 450 & 50 decreased \\
$\epsilon_{90}$ (m) & 825 & 600 & 47 decreased \\
\midrule
$S_1/N$ at 400 m (\%) & 22.1 & 73.6 & 49 increased \\
Largest-jump magnitude
(percentage points) & 25.7 & 18.2 & 40 decreased \\
\bottomrule
\end{tabular}
\end{table}

% Figure 2(a) — First page
\begin{figure*}[p]
\centering

\begin{subfigure}[t]{0.85\textwidth}
    \centering
    \includegraphics[
        width=\linewidth,
        height=0.75\textheight,
        keepaspectratio
    ]{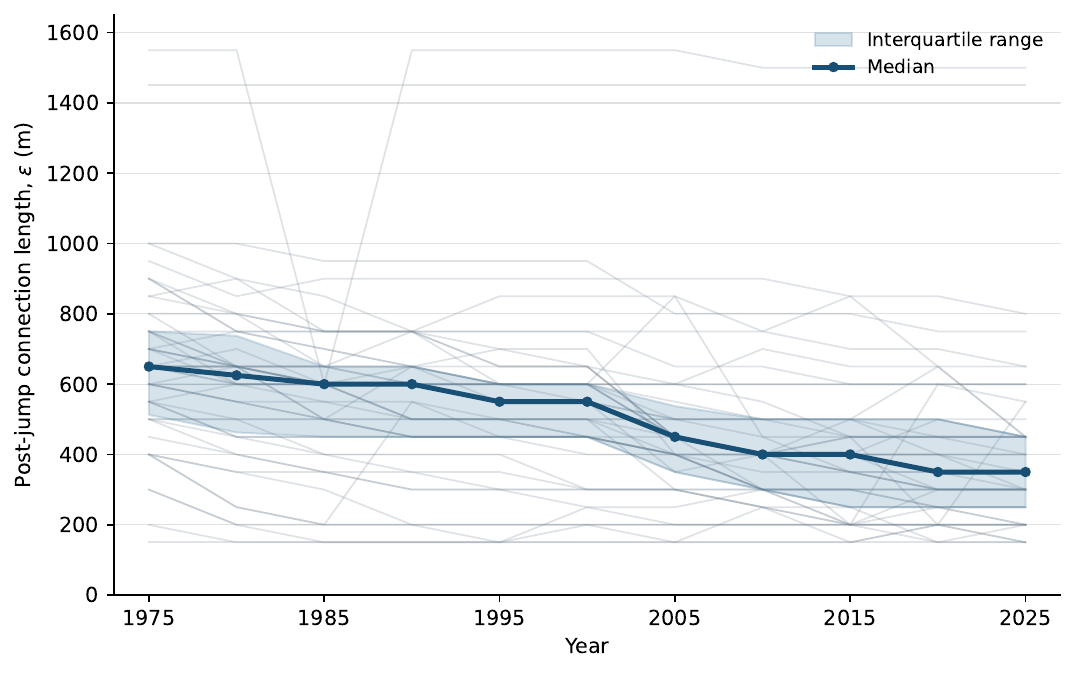}
    \caption{Evolution of characteristic connection
    lengths, 1975--2025}
\end{subfigure}

\caption{
\textbf{Percolation reveals a systematic reduction
in characteristic spatial connection lengths
across India's 50 largest urban agglomerations.}
(a) Post-jump DBSCAN connection lengths across
eleven observations from 1975 to 2025.
Thin gray lines represent individual cities;
the solid blue line indicates the median,
and the shaded band represents the interquartile range.
}
\label{fig:shrinking_length}

\end{figure*}

\clearpage

% Figure 2(b) — Second page
\begin{figure*}[p]
\ContinuedFloat
\centering

\begin{subfigure}[t]{0.95\textwidth}
    \centering
    \includegraphics[
        width=\linewidth,
        height=0.82\textheight,
        keepaspectratio
    ]{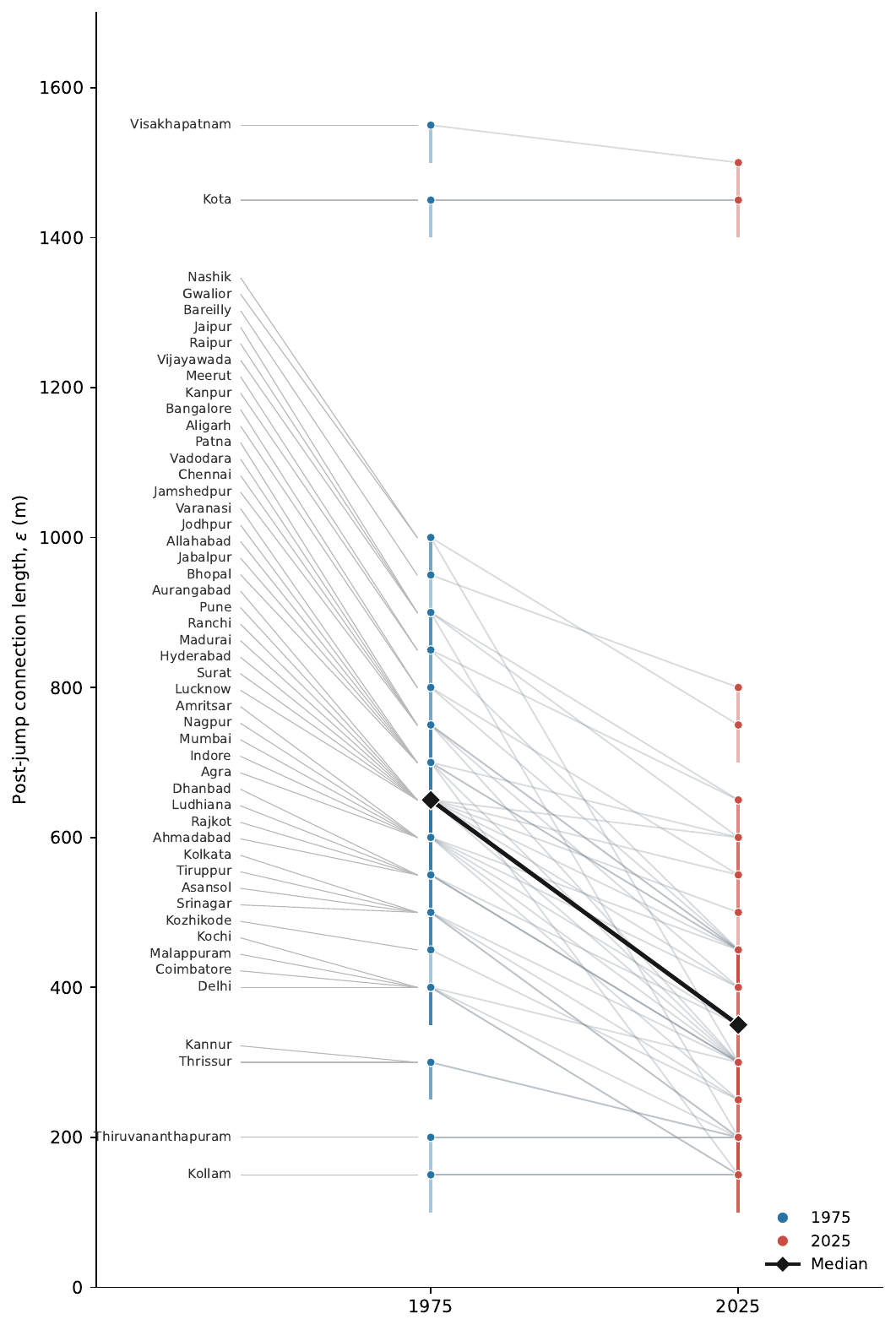}
    \caption{Changes in characteristic connection
    lengths across 50 cities}
\end{subfigure}

\caption{
\textbf{Percolation reveals a systematic reduction
in characteristic spatial connection lengths
(continued).}
(b) Paired comparison of connection lengths
in 1975 and 2025. Lines connect observations
for the same city. Blue and red markers
identify 1975 and 2025, respectively.
Of the 50 cities, 47 exhibit shorter
connection lengths, three remain unchanged,
and none exhibits an increase.
}

\end{figure*}

\clearpage

\subsection{Percolation reveals distinct morphologies of urban coalescence}

The largest connectivity transitions reveal distinct spatial morphologies through which urban settlement systems coalesce. Figure~\ref{fig:morphologies} shows the ten largest connected components immediately before and after the largest observed merger for four representative cities in 2025, illustrating different configurations of spatial integration.

Delhi illustrates \textbf{compact consolidation} (Figure~\ref{fig:morphologies}(a,b)). At $\epsilon=250$ m, its dominant urban core already contains 75.3\% of eligible settlement cells. Increasing $\epsilon$ to 300 m raises this share to 83.0\%, indicating relatively limited absorption of peripheral settlement components into an already dominant urban core.

Mumbai exhibits \textbf{corridor coalescence} (Figure~\ref{fig:morphologies}(c,d)). Its elongated north--south settlement structure reflects its constrained coastal geography. Increasing $\epsilon$ from 300 to 350 m raises the largest-cluster share from 35.8\% to 57.6\%, demonstrating how connectivity emerges through the linking of settlement corridors rather than radial consolidation around an urban core.

Bhopal illustrates \textbf{satellite absorption} (Figure~\ref{fig:morphologies}(e,f)). At $\epsilon=550$ m, a dominant urban core remains separated from several smaller peripheral settlements. Increasing $\epsilon$ to 600 m incorporates multiple secondary components, raising the largest-cluster share from 73.7\% to 88.1\% and extending the connected metropolitan footprint.

Rajkot represents \textbf{regional multi-cluster coalescence} (Figure~\ref{fig:morphologies}(g,h)). At $\epsilon=250$ m, several substantial settlement components remain disconnected, with the largest containing only 21.0\% of eligible cells. Increasing $\epsilon$ to 300 m produces a pronounced merger, raising this share to 78.9\% and creating an extensive connected regional settlement system.

These configurations are illustrative rather than mutually exclusive: individual urban systems may combine corridor development, satellite absorption, compact consolidation, and regional coalescence. Together with the longitudinal evidence of shrinking characteristic connection lengths, these morphologies reveal that increasing settlement connectivity can accompany distinct spatial configurations of urban growth, rather than a uniform process of compact consolidation. These differences are consequential for SDG 11.3.1, since alternative configurations of spatial integration can alter the relationships between population growth, built-up surface area, and urban footprint expansion. Percolation thus reveals a morphological mechanism of land-consumptive growth that aggregate measures of urban expansion alone cannot capture.

%These configurations are illustrative rather than mutually exclusive: individual urban systems may combine corridor development, satellite absorption, compact consolidation, and regional coalescence. The distinction is consequential for SDG 11.3.1, since different forms of spatial integration may produce different relationships between population growth, built-up surface area, and urban footprint expansion. Percolation therefore reveals not only the extent of connected settlement systems, but also the heterogeneous spatial structures underlying their formation.

% FIGURE 3 — PAGE 1
% Delhi and Mumbai

\begin{figure*}[p]
\centering

% Row 1: Compact consolidation - Delhi

\begin{subfigure}[t]{0.48\textwidth}
    \centering
    \includegraphics[
        width=\linewidth,
        height=0.4\textheight,
        keepaspectratio
    ]{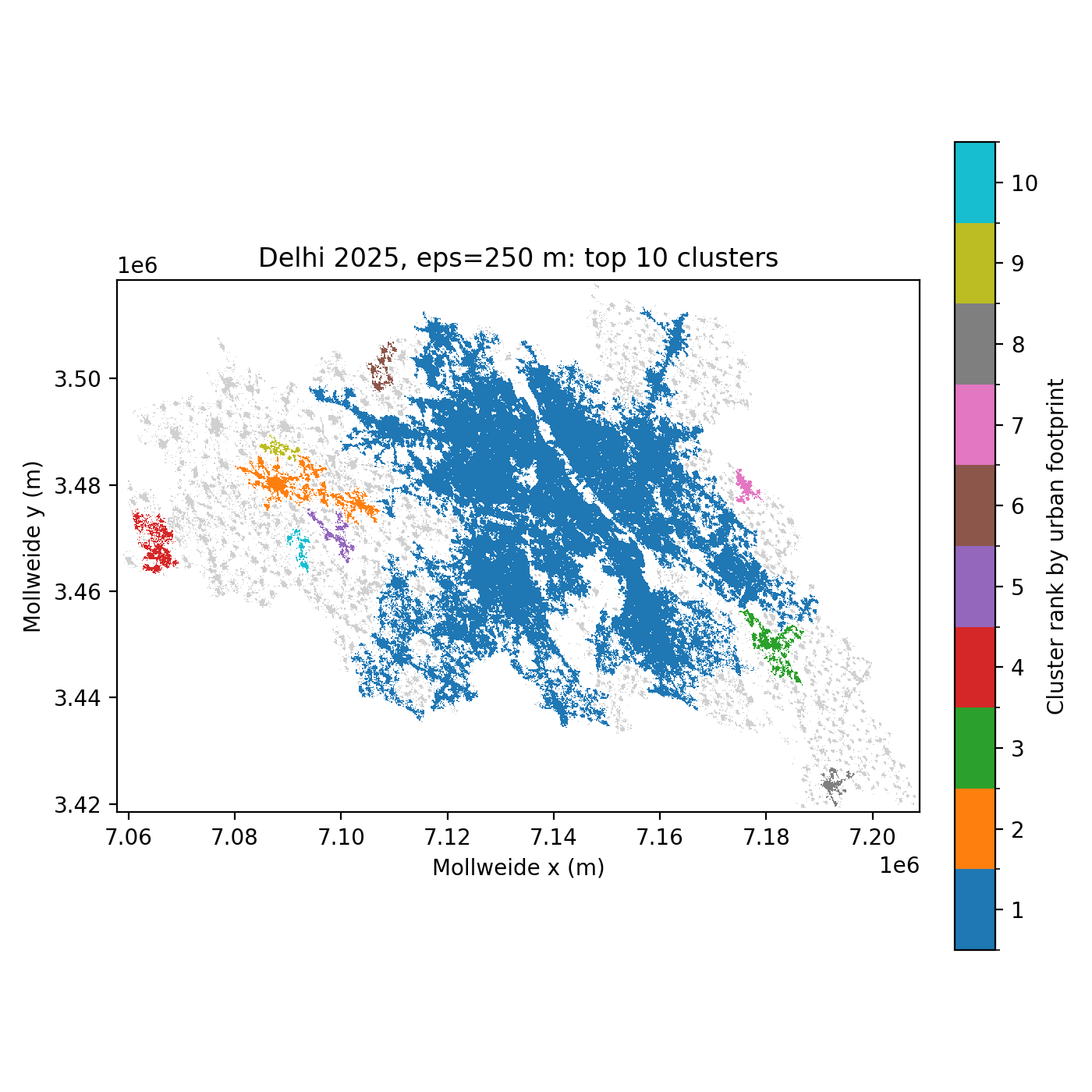}
    \caption{Delhi, pre-jump ($\epsilon=250$~m)}
\end{subfigure}
\hfill
\begin{subfigure}[t]{0.48\textwidth}
    \centering
    \includegraphics[
        width=\linewidth,
        height=0.4\textheight,
        keepaspectratio
    ]{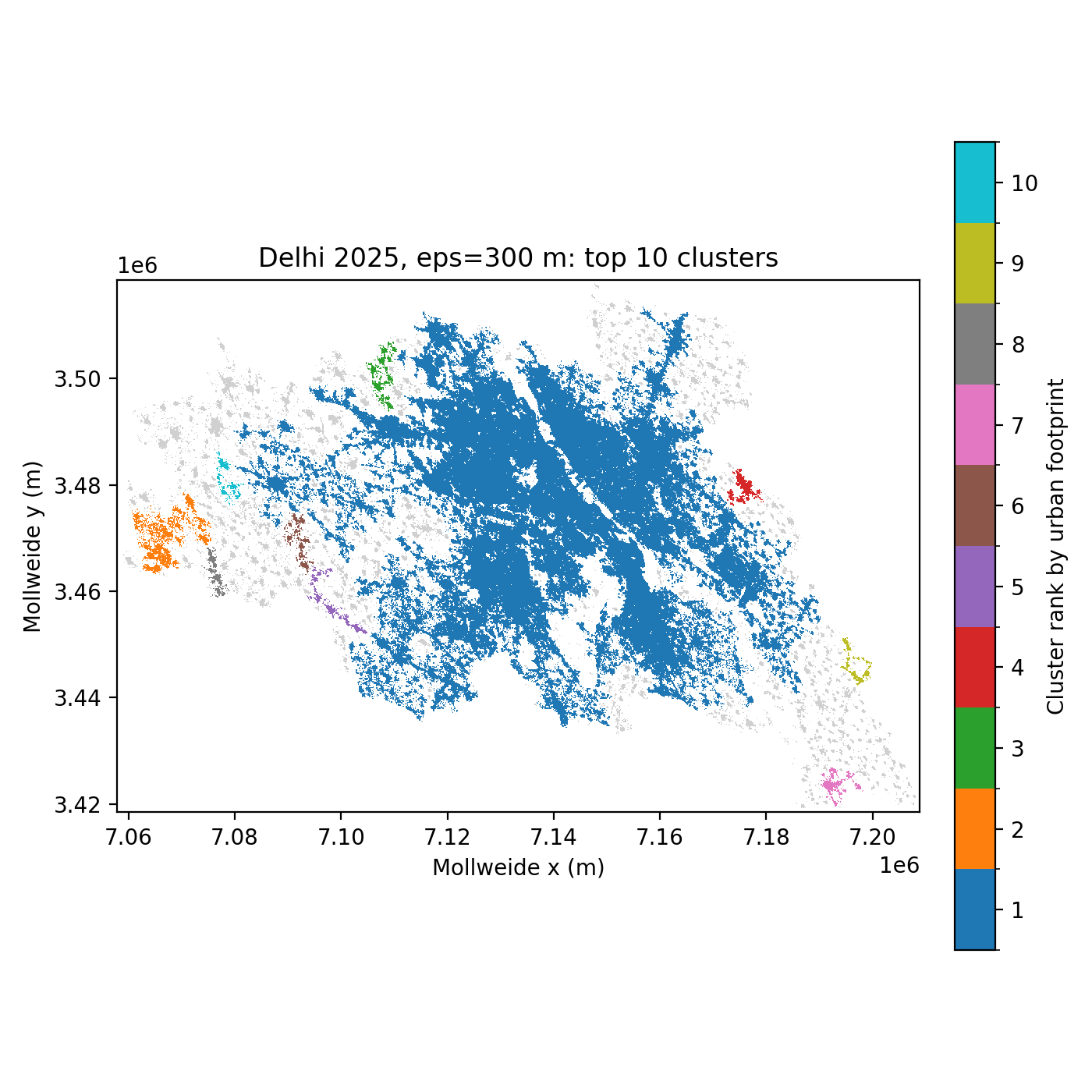}
    \caption{Delhi, post-jump ($\epsilon=300$~m)}
\end{subfigure}

\vspace{0.8em}

% Row 2: Corridor coalescence - Mumbai

\begin{subfigure}[t]{0.48\textwidth}
    \centering
    \includegraphics[
        width=\linewidth,
        height=0.4\textheight,
        keepaspectratio
    ]{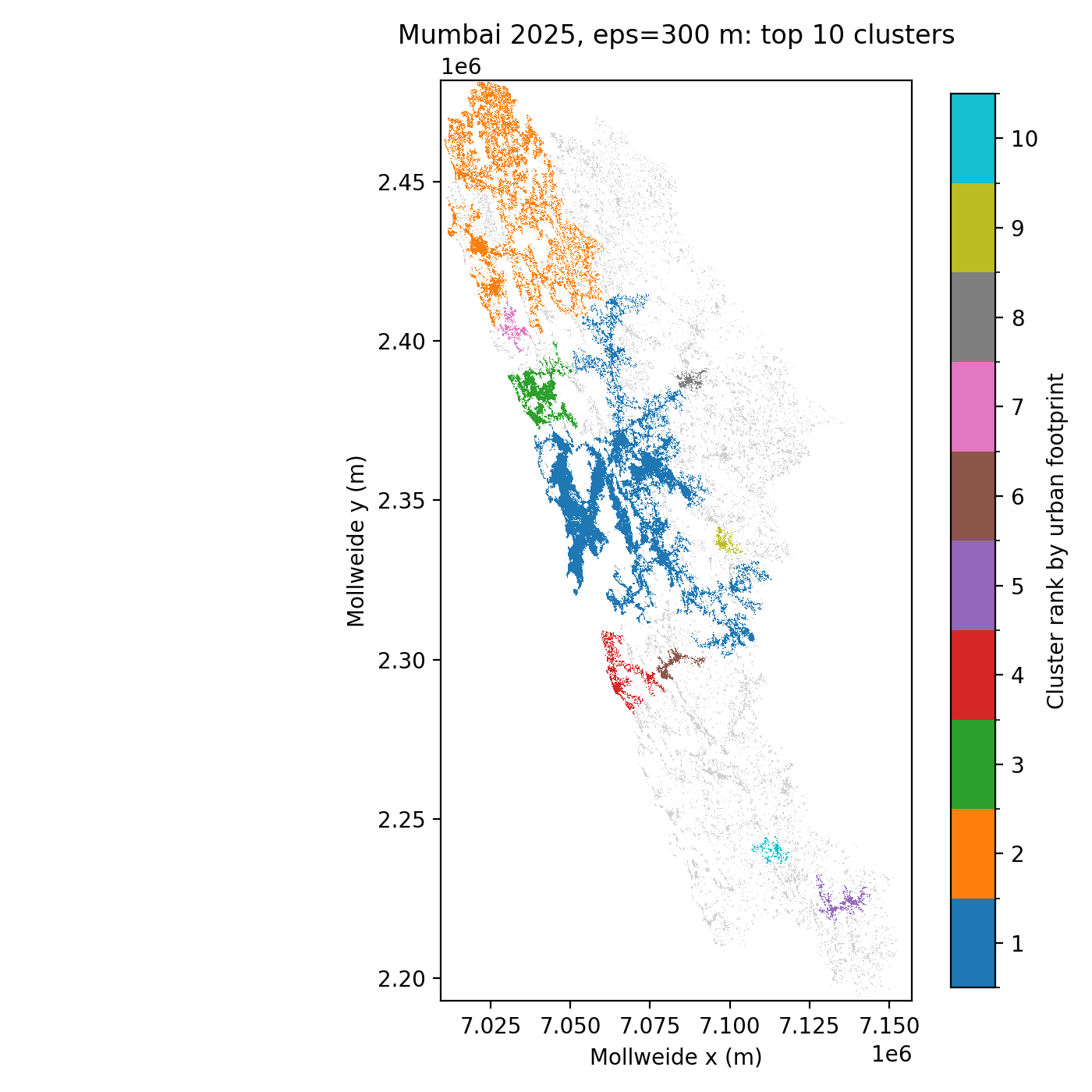}
    \caption{Mumbai, pre-jump ($\epsilon=300$~m)}
\end{subfigure}
\hfill
\begin{subfigure}[t]{0.48\textwidth}
    \centering
    \includegraphics[
        width=\linewidth,
        height=0.4\textheight,
        keepaspectratio
    ]{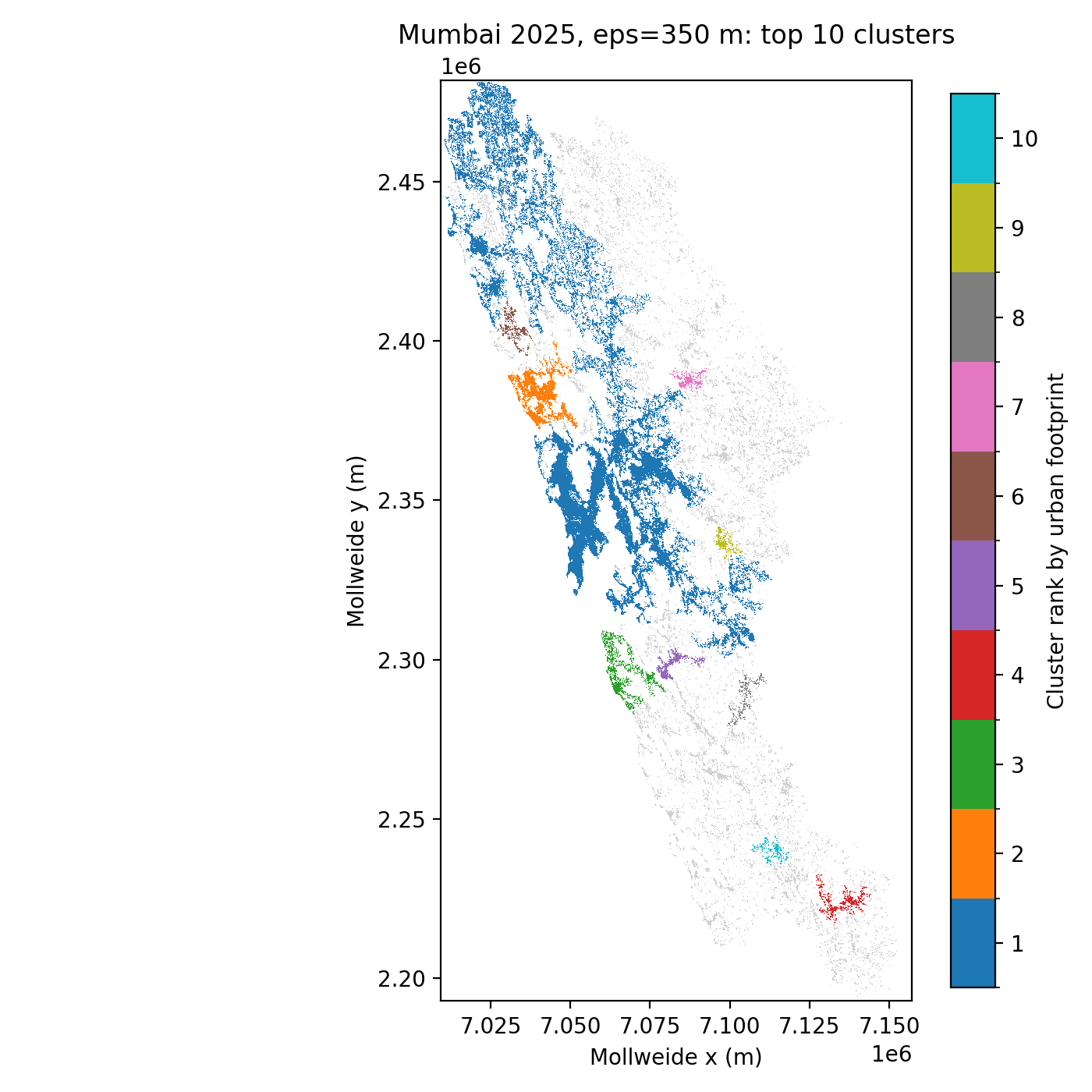}
    \caption{Mumbai, post-jump ($\epsilon=350$~m)}
\end{subfigure}

\vspace{0.5em}

\centering
\small\textit{Figure continued on the following page.}

\end{figure*}

\clearpage

% FIGURE 3 — PAGE 2
% Bhopal and Rajkot

\begin{figure*}[p]
\ContinuedFloat
\centering

% Continue subfigure lettering from (e)
\setcounter{subfigure}{4}

% Row 3: Satellite absorption - Bhopal

\begin{subfigure}[t]{0.48\textwidth}
    \centering
    \includegraphics[
        width=\linewidth,
        height=0.4\textheight,
        keepaspectratio
    ]{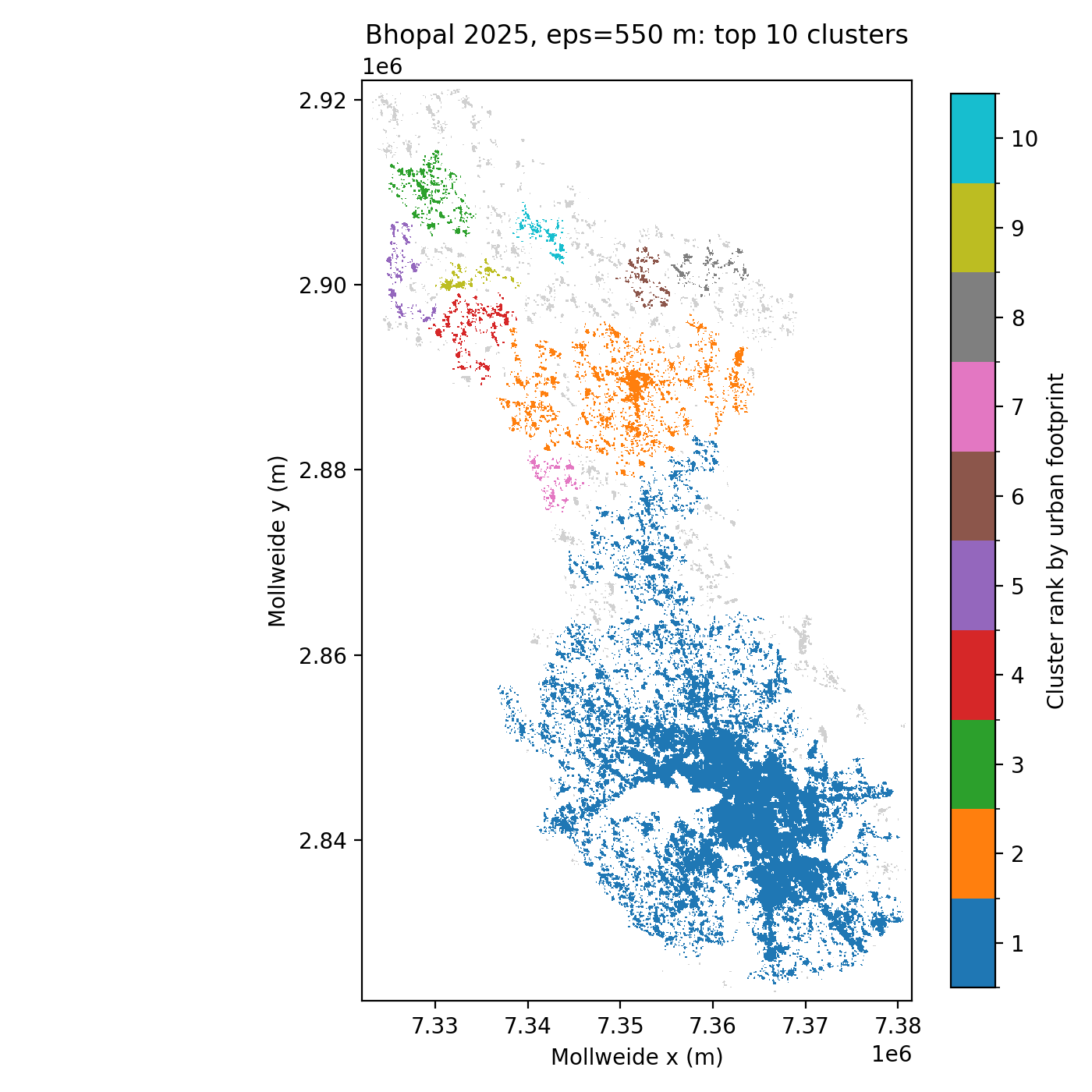}
    \caption{Bhopal, pre-jump ($\epsilon=550$~m)}
\end{subfigure}
\hfill
\begin{subfigure}[t]{0.48\textwidth}
    \centering
    \includegraphics[
        width=\linewidth,
        height=0.4\textheight,
        keepaspectratio
    ]{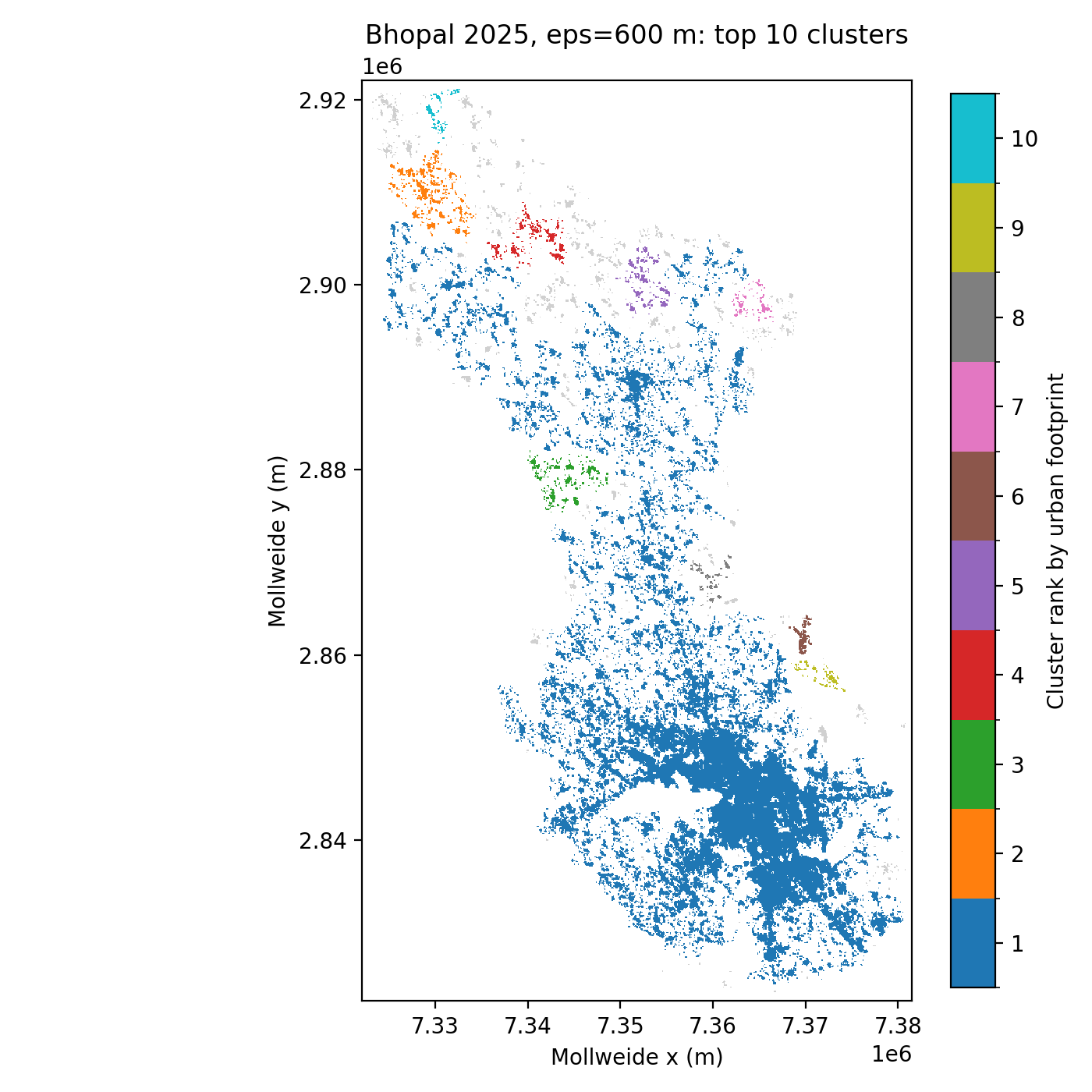}
    \caption{Bhopal, post-jump ($\epsilon=600$~m)}
\end{subfigure}

\vspace{0.8em}

% Row 4: Regional coalescence - Rajkot

\begin{subfigure}[t]{0.48\textwidth}
    \centering
    \includegraphics[
        width=\linewidth,
        height=0.4\textheight,
        keepaspectratio
    ]{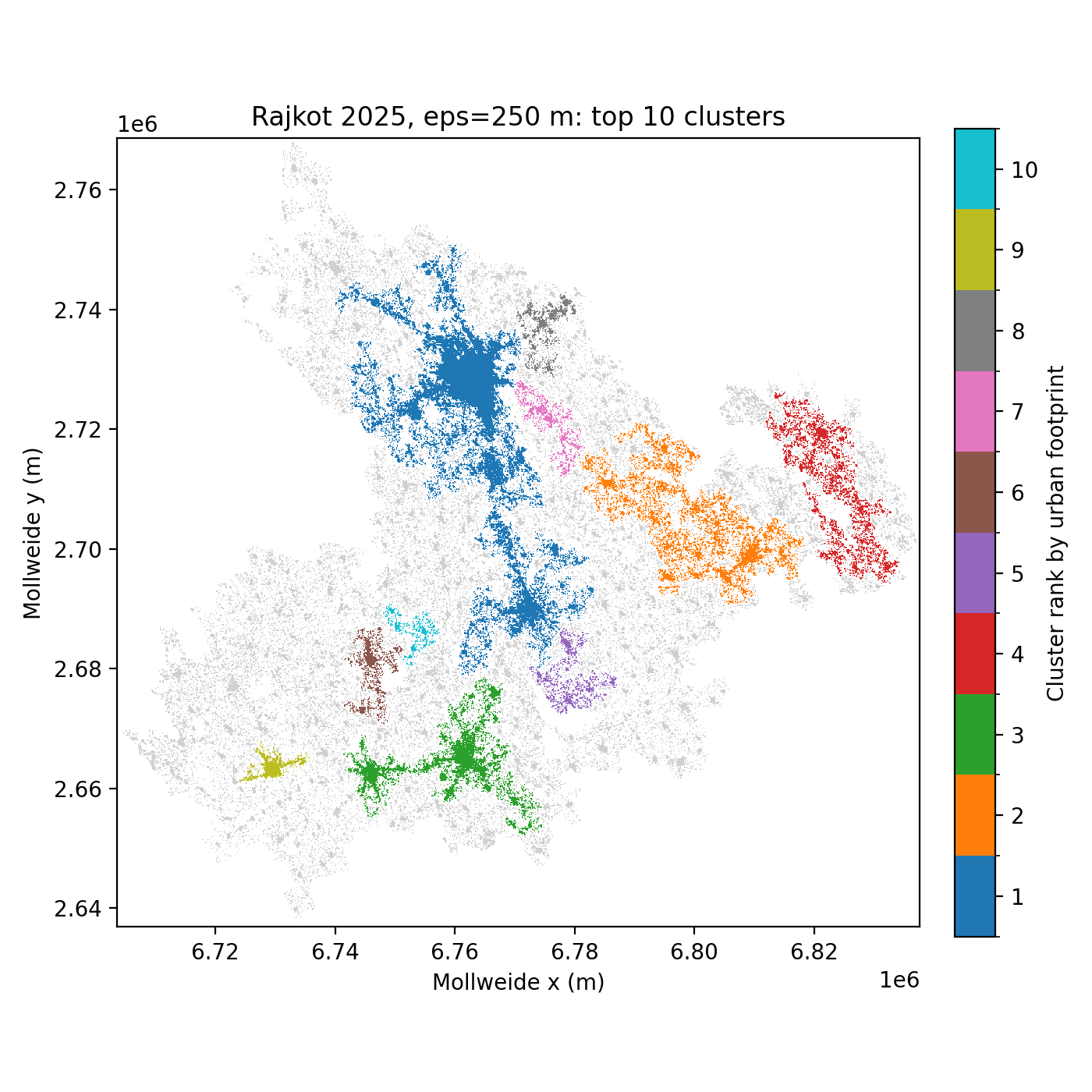}
    \caption{Rajkot, pre-jump ($\epsilon=250$~m)}
\end{subfigure}
\hfill
\begin{subfigure}[t]{0.48\textwidth}
    \centering
    \includegraphics[
        width=\linewidth,
        height=0.4\textheight,
        keepaspectratio
    ]{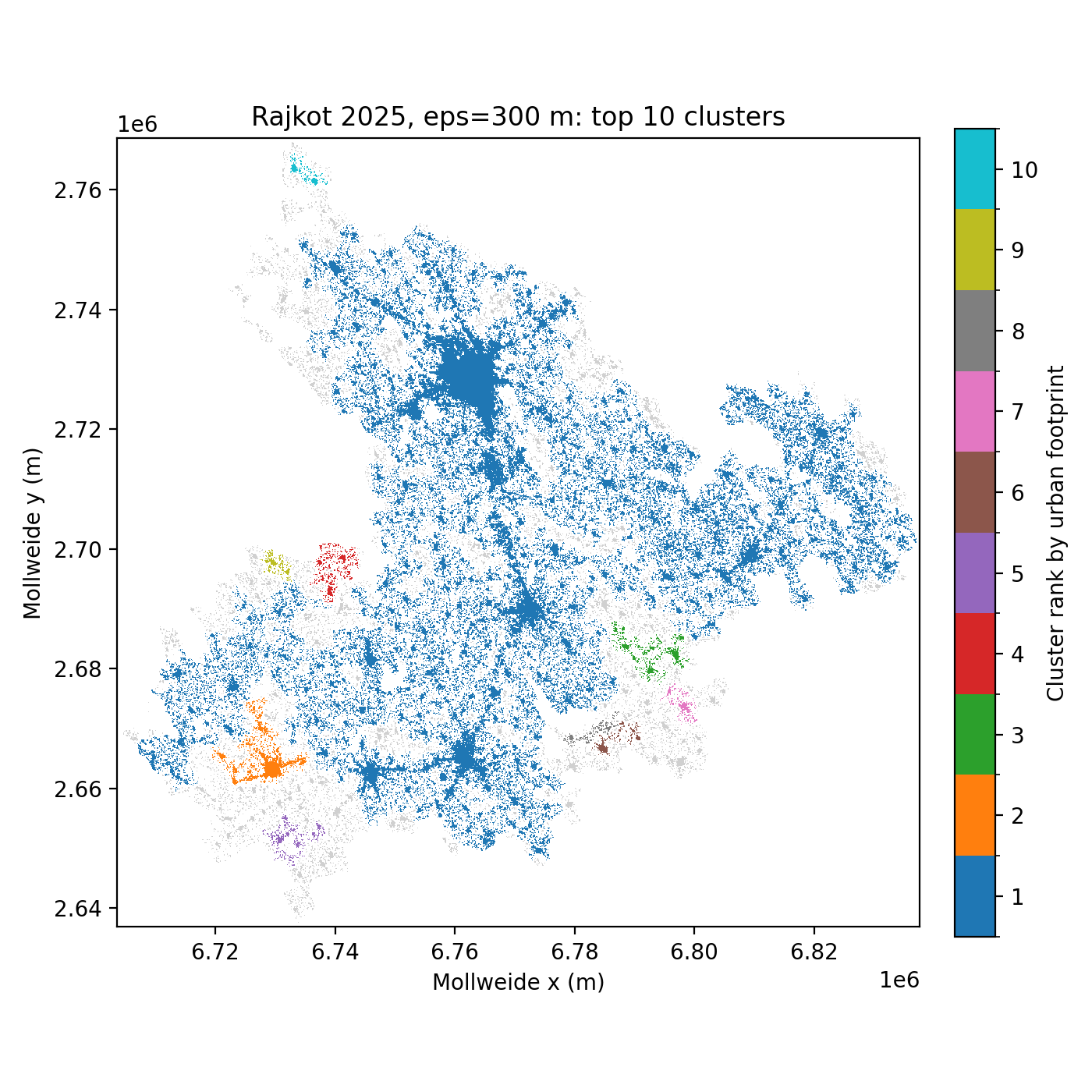}
    \caption{Rajkot, post-jump ($\epsilon=300$~m)}
\end{subfigure}

\caption{
\textbf{Distinct morphological configurations of
urban settlement coalescence revealed by percolation.}
The ten largest connected settlement clusters
immediately before (left column) and after
(right column) the largest observed increase
in $S_1/N$ for four representative Indian
cities in 2025.
(a,b) Compact consolidation in Delhi:
an already established dominant urban core
undergoes relatively limited peripheral
integration without substantial reconfiguration
of its connected footprint.
(c,d) Corridor coalescence in Mumbai:
spatial connectivity increases along an
elongated north--south settlement configuration.
(e,f) Satellite absorption in Bhopal:
multiple secondary peripheral settlements
become incorporated into the dominant urban
cluster.
(g,h) Regional multi-cluster coalescence
in Rajkot: several substantial disconnected
settlement components merge into a larger
regional urban system.
Colors distinguish clusters by size rank
within each panel and do not represent
persistent cluster identities across
connection lengths. All maps use
unrestricted DBSCAN ($\mathrm{minPts}=1$).
The four examples illustrate potentially
overlapping morphological processes rather
than mutually exclusive urban categories.
}
\label{fig:morphologies}

\end{figure*}

\clearpage

\subsection{Urban delineation reverses the relative growth of urban footprints and built-up surfaces}

\begin{figure}[h]
\centering
\includegraphics[width=1.0\textwidth]{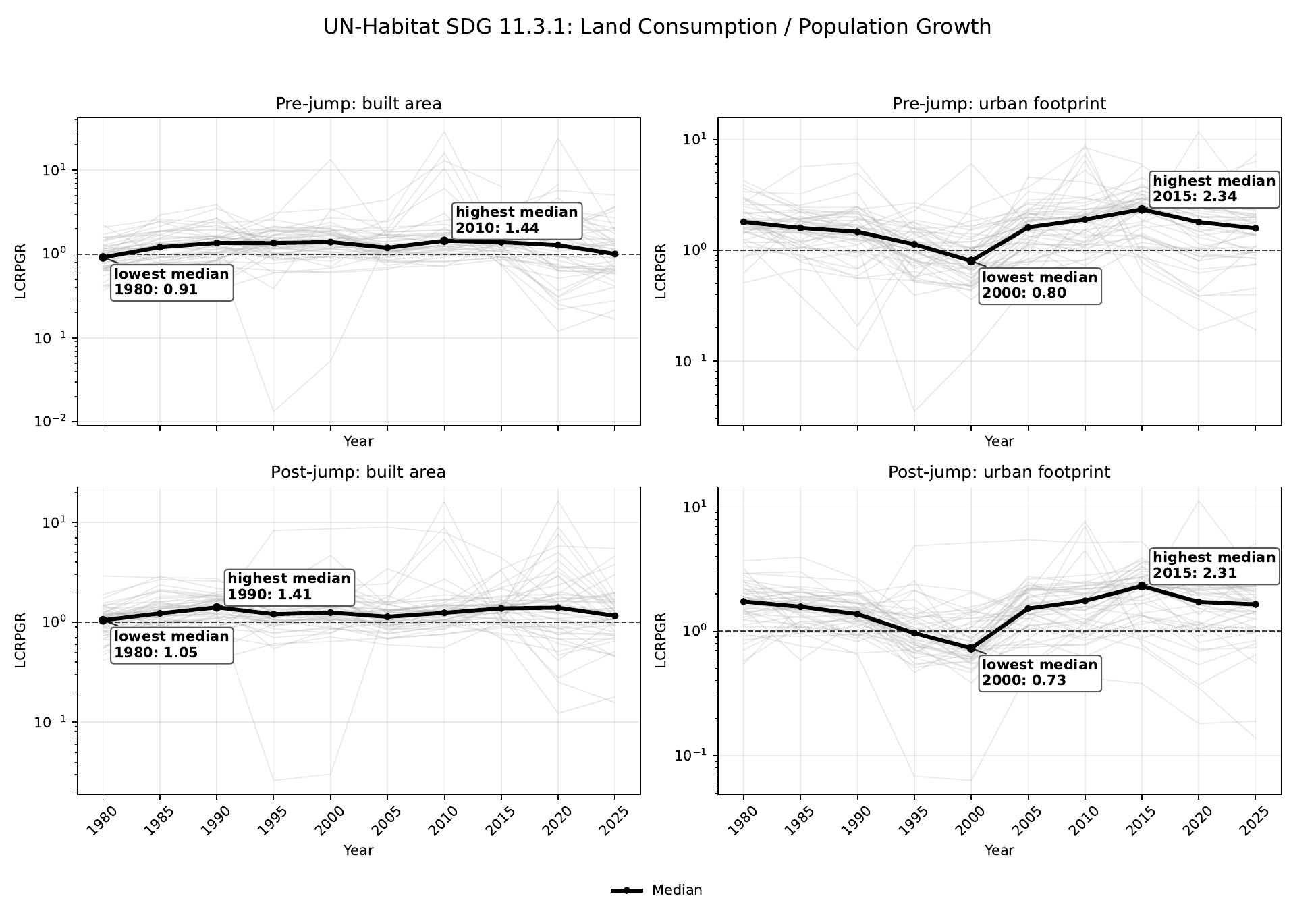}
\caption{\textbf{Land consumption relative to population growth depends on urban footprint definition.}
Land consumption rate to population growth rate ratios (LCRPGR) for 50 Indian urban agglomerations across ten five-year intervals from 1980 to 2025. The four panels show (a) pre-jump built-up surface area (top left), (b) pre-jump urban footprint (top right), (c) post-jump built-up surface area (bottom left), and (d) post-jump urban footprint (bottom right). Thin gray lines represent individual city trajectories, while solid black lines indicate annual cross-sectional medians. The vertical axes use logarithmic scales, and horizontal dashed lines mark LCRPGR = 1, indicating proportional land and population growth when both growth rates are positive. Values above unity indicate land consumption outpacing population growth under the same condition. Annotations identify the years and values of the highest and lowest annual medians in each panel. Nonpositive and undefined ratios are excluded from the logarithmic plots.}\label{fig:lcrpgr}
\end{figure}

Land consumption relative to population growth depends strongly on how urban land is defined. We estimate SDG Indicator 11.3.1, the ratio of land consumption rate to population growth rate (LCRPGR), for pre- and post-jump clusters using two measures: built-up surface area and urban footprint, defined as the area of occupied 100 m grid cells (see Methods). For positive population and land-growth rates, ratios above one indicate that land consumption outpaces population growth.

Figure~\ref{fig:lcrpgr} shows LCRPGR trajectories across 50 cities from 1980 to 2025. Urban-footprint-based ratios reveal two periods during which land consumption was particularly high relative to population growth: an early phase during 1980--1990, followed by renewed relative expansion after 2005. Between these periods, median ratios decline to 0.80 for pre-jump and 0.73 for post-jump footprints in 2000. Both subsequently reach their highest median values in 2015, at 2.34 and 2.31, respectively. These results indicate that the spatial extent of connected settlements frequently expands faster than their populations.

Built-area LCRPGR exhibits a less pronounced temporal pattern, with ratios generally closer to one. The pre-jump median ranges from 0.91 in 1980 to a peak of 1.44 in 2010, while the post-jump median ranges from 1.05 in 1980 to a peak of 1.41 in 1990. Across all positive, finite observations, median built-area ratios are 1.25 and 1.24 for pre- and post-jump clusters, respectively, compared with 1.52 and 1.48 for urban footprints. These longitudinal patterns of land-consumptive growth coexist with sublinear cross-sectional scaling of built-up surface and urban footprint with population (Supplementary Fig.~\ref{fig:urban_scaling}(b--e)), emphasizing that relationships across cities do not necessarily describe growth trajectories within individual cities~\cite{MarquisBarthelemy2026}.

The difference between these measures captures distinct dimensions of urban land development. Urban footprint captures the spatial extent of occupied settlement cells, including incompletely built-up areas, whereas built-up surface measures the amount of built land within those cells. The higher footprint-based ratios are consistent with peripheral spatial expansion occurring alongside development within existing settlement areas, although these aggregate measures do not separately identify infill and outward growth.

Across the fixed district-based study regions, median eligible urban footprint increased from 447.6 km$^2$ in 1975 to
1,382.7 km$^2$ in 2025 (Supplementary Fig.~1). Population densities within percolation-defined clusters also generally declined over this period. Measured per unit of eligible footprint, median pre-jump density decreased from 8,172 to 5,456 people/km$^2$, while post-jump density decreased from 5,935 to 4,634 people/km$^2$ (Supplementary Fig.~3). These longer-term density declines are consistent with the land-consumptive growth identified through LCRPGR,
although the density trajectories use independently determined cluster boundaries in each observation year,
whereas LCRPGR comparisons hold the connection length constant within each five-year interval.

Over the full 1975--2025 period, cumulative land growth reveals an even stronger dependence on urban delineation. Across the 50 cities, the summed urban footprint of pre-jump clusters increased by 386\%, exceeding the 361\% increase in built-up surface area. For post-jump clusters, however, this relationship reversed: built-up surface increased by 315\%, compared with 287\% for urban footprint. This reversal demonstrates that connectivity-based delineation does not merely alter the magnitude of measured land consumption, but can fundamentally change the apparent balance between spatial expansion and development within occupied settlement areas. The same underlying settlement data can therefore yield different interpretations of urban growth depending on which connected components are included, illustrating a consequential manifestation of the MAUP in urban sustainability measurement.

These findings connect the morphological transformation of settlement systems to the measurement of land-consumptive growth. As characteristic connection lengths shrink, spatially separated settlements become integrated into larger connected systems through the distinct coalescence configurations identified above. The incorporation of surrounding settlements can alter the relative growth of population, built-up surface, and urban footprint, producing either higher or lower LCRPGR values than those measured for the corresponding pre-jump clusters. Thus, understanding how settlement components merge is essential not only for characterizing the spatial organization of urban growth, but also for interpreting how much land is being consumed relative to population growth under SDG 11.3.1 and how.

%Over the full 1975--2025 period, cumulative land growth reveals
%an even stronger dependence on urban delineation. Across the 50
%cities, the summed urban footprint of pre-jump clusters increased
%by 386\%, exceeding the 361\% increase in built-up surface area.
%For post-jump clusters, however, this relationship reversed:
%built-up surface increased by 315\%, compared with 287\% for
%urban footprint. These findings
%demonstrate that urban delineation does not merely influence
%estimates of land consumption: it can reverse conclusions about
%whether urban growth is dominated by spatial expansion or
%built-up intensification. 

%The pre- and post-jump comparisons further demonstrate that sustainability estimates depend on the connection length used to delineate settlements. Broader post-jump clusters may exhibit either higher or lower LCRPGR values than their corresponding pre-jump cores, depending on the growth characteristics of incorporated settlements. Thus, both the definition of urban land and the spatial extent of the connected system influence assessments of land-consumptive growth under SDG 11.3.1.

\subsection{Land-consumptive growth and settlement connectivity contraction show little dependence on city size}

\begin{figure}[p]
\centering

\includegraphics[width=0.82\textwidth]{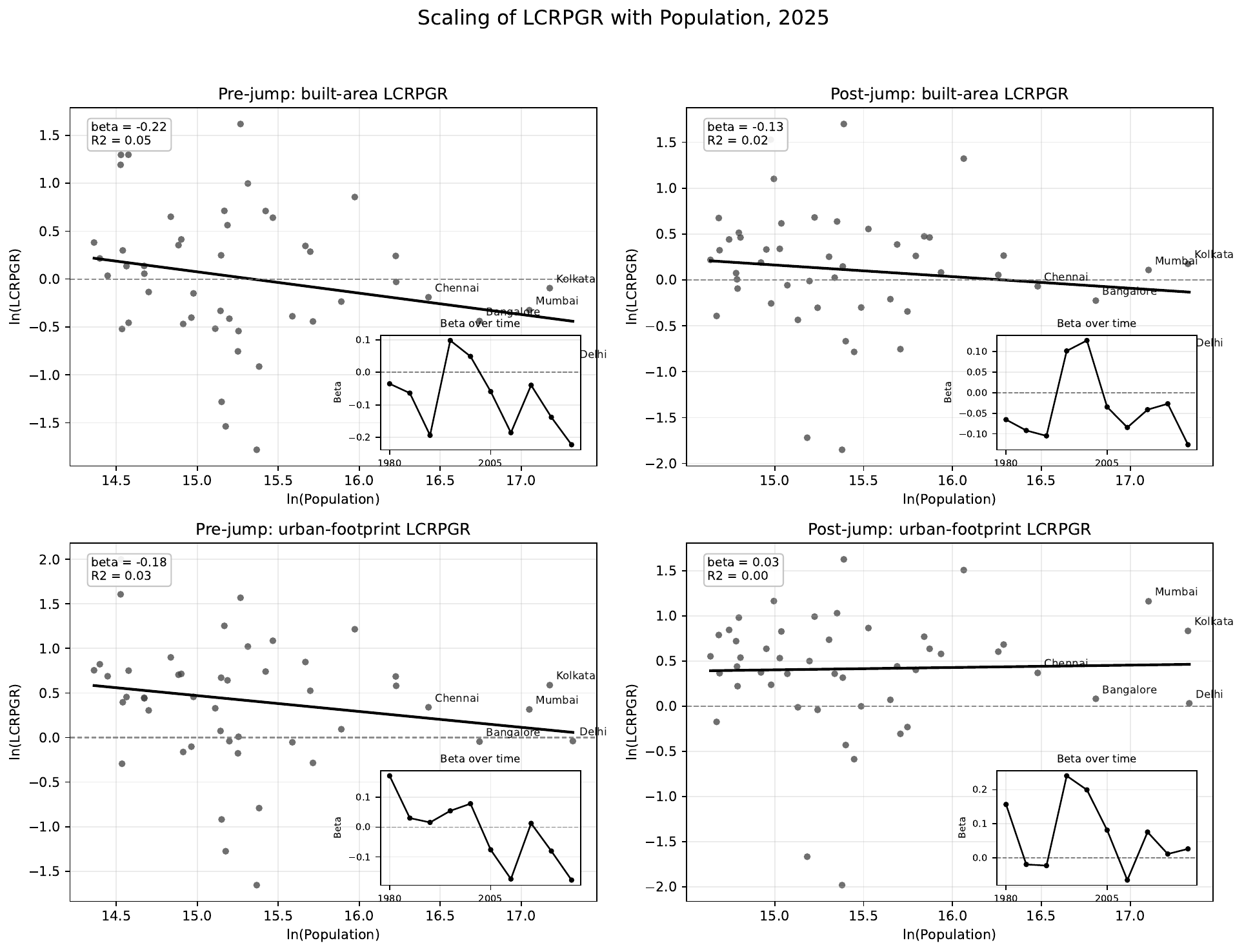}

\medskip

\includegraphics[width=0.50\textwidth]{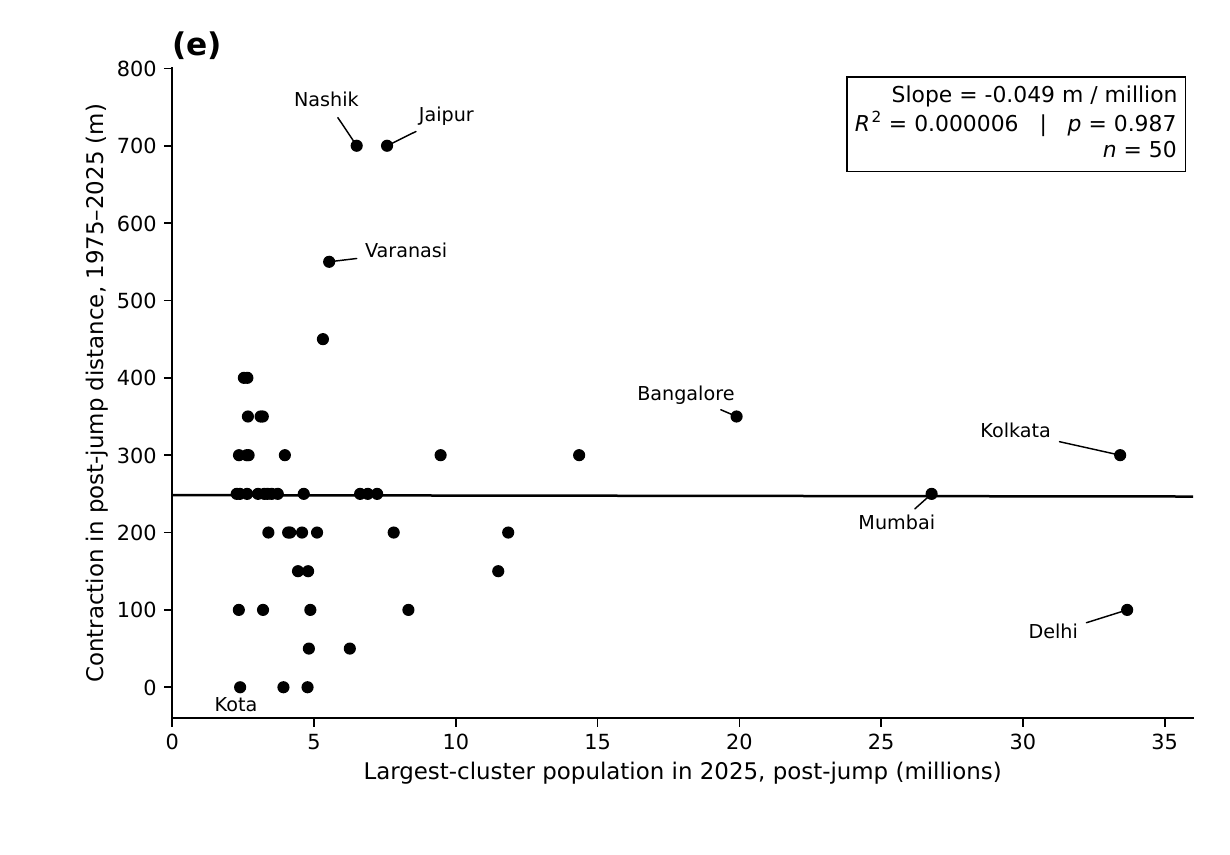}

\caption{\textbf{Land-consumptive growth and contraction in settlement connectivity show weak relationships with city size.}
Panels (a)--(d) show cross-sectional relationships between the natural logarithm of LCRPGR and the natural logarithm of cluster population across Indian urban agglomerations in 2025: (a) pre-jump built-up surface area, (b) post-jump built-up surface area, (c) pre-jump urban footprint, and (d) post-jump urban footprint. Gray points represent individual cities, and solid black lines indicate fitted log-linear regressions. The estimated scaling exponent $\beta$ and coefficient of determination $R^2$ are reported in each panel. Horizontal dashed lines indicate $\ln(\mathrm{LCRPGR})=0$, corresponding to LCRPGR $=1$. Insets show the temporal evolution of the fitted exponent $\beta$ across successive five-year observations from 1980 to 2025, with dashed lines indicating $\beta=0$. Only positive, finite LCRPGR observations are included in these regressions. Panel (e) shows the relationship between 2025 post-jump largest-cluster population and the contraction in characteristic post-jump connection length between 1975 and 2025. Black points represent individual cities, and the solid black line indicates the fitted ordinary least-squares regression. The slope, coefficient of determination ($R^2$), and $p$-value are reported in the panel.}

\label{fig:lcrpgr_scaling}
\end{figure}

%Comment BEGIN UNSDG-REVIEW-08
% City-size claim, slope uncertainty, and excluded observations (Section 2.6, Figure 6, Methods 4.5–4.6).
%
% Suggested replacement Section 2.6 heading: 'Land-consumption ratios show weak association with city size'.
%
% Suggested Results wording: '
%Within the positive, finite LCRPGR observations retained for the log-linearmodels, population explains little of the variation in land-consumption ratios. The fitted relationships provide limited predictive power; they do not establish that the population association is exactly zero.
%
% Please report confidence intervals for slopes and the sample size of each fit. Supply counts of excluded
% observations by year and delineation, distinguishing zero population growth, negative ratios, and positive
% ratios generated by simultaneous land and population decline. Report whether the last category enters the
% displayed medians and regressions; it should not be interpreted as land-consumptive growth. Very small
% population-growth denominators can also generate large ratios, so examine their influence before making a
% broad null claim.
%
% Retain the existing UN-Habitat metadata citation \cite{UNHabitat2020Metadata1131} for the indicator formula
% and interpretation. These statistical qualifications concern the present analysis and require reported
% uncertainty/sample accounting, rather than an added citation. No uncertainty estimates or sensitivity
% results should be asserted before calculation.
%Comment END UNSDG-REVIEW-08

We finally examine whether city size is systematically associated with two distinct features of urban transformation: land consumption relative to population growth and the contraction of characteristic settlement connection lengths. Across both built-up surface and urban footprint measures, the relationship between population and LCRPGR is weak, with fitted exponents close to zero and low explanatory power (Figure~\ref{fig:lcrpgr_scaling}). In 2025, built-area LCRPGR exhibits exponents of $\beta=-0.222$ ($R^2=0.053$) for pre-jump clusters and $\beta=-0.126$ ($R^2=0.017$) for post-jump clusters. The corresponding urban-footprint exponents are $\beta=-0.178$ ($R^2=0.033$) and $\beta=0.026$ ($R^2=0.001$), respectively. These regressions include only observations with positive, finite LCRPGR values.

The contraction in characteristic connection length is similarly unrelated to population size in a linear regression. Figure~\ref{fig:lcrpgr_scaling}(c) relates the 1975--2025 reduction in post-jump merger distance to the population of the largest post-jump cluster in 2025. Across 50 cities, the fitted slope is $-0.049$ m per million people ($R^2=0.000006$, $p=0.987$), providing no detectable linear relationship between city size and the magnitude of contraction. The dispersion, however, is not uniform across the observed urban hierarchy. Intermediate-sized agglomerations (5--15 million people) exhibit contractions ranging from 50 to 700 m, compared with 100 to 350 m for those above 15 million. This difference is descriptive: only four agglomerations exceed 15 million people, limiting inferences about how variability changes with size.

These findings concern distinct dimensions of urban growth. LCRPGR describes the rate of land expansion relative to population growth, whereas the contraction in characteristic connection lengths describes the changing spatial organization of settlement clusters. Neither is strongly predicted by population size in the fitted relationships. The distinct coalescence morphologies identified above therefore provide complementary information on how settlement systems expand and integrate, beyond what can be inferred from city size or aggregate land-consumption measures alone.

\section{Discussion and Conclusions}

Shrinking characteristic settlement-merger distances reveal a morphological mechanism of land-consumptive urbanization that aggregate expansion measures overlook. Greater connectivity does not necessarily imply compactness, rising density, or reduced land consumption: compact, corridor, satellite, and regional configurations represent different pathways of coalescence. Neither contraction magnitude nor land-consumption ratios show strong associations with city size. Cross-sectional economies of land use likewise need not translate into more land-efficient longitudinal growth~\cite{MarquisBarthelemy2026}.

Urban delineation affects not merely the estimated magnitude but the apparent character of land consumption. The reversal between pre- and post-merger comparisons illustrates the modifiable areal unit problem (MAUP): different connected units yield different assessments of footprint expansion relative to built-up coverage. For SDG 11.3.1, both the land measure and the delineation of the city matter. Percolation makes this sensitivity visible without eliminating dependence on raster resolution, cell-eligibility criteria, connection lengths, or study-region boundaries.

The largest merger is an operational transition in a finite settlement system, not a uniquely defined city boundary: infact it illustrates the fundamental problem in the very concept of a unique city boundary. Because clusters are defined independently each year, observed changes combine new development with the incorporation of existing settlements. The declining merger distance thus illuminates a morphological process shaping  land-consumptive growth.

Planning attention should focus on the gaps between growing settlements. Transport investment, infrastructure provision, and land-use regulation can shape whether these become appropriate infill locations or undergo damaging conversion. Coalescence is not inherently sustainable: closing gaps and simultaneous expansion consume agricultural land, fragment ecosystems, or increase exposure to hazards. Combining connectivity analysis with transport networks, land-use regulations, agricultural and ecological assets, and environmental risks could guide infrastructure investment, prioritize suitable infill, protect sensitive landscapes, and support planning across administrative boundaries. Sustainable urbanization requires understanding not only how much land is consumed, but where and how settlements grow together.

\section{Methods}\label{methods}

\subsection{Data and study regions}

We use the GHS-POP and GHS-BUILT-S products from the Global Human Settlement Layer (GHSL) to reconstruct urban settlement patterns across India's 50 largest urban agglomerations at five-year intervals between 1975 and 2025~\cite{schiavina_etal_2023_ghs_pop,pesaresi_politis_2023_ghs_built_s}. The GHSL R2023A series combines historical estimates with projections for 2025; accordingly, the 2025 results describe modeled settlement configurations rather than independently observed conditions. GHS-POP provides population counts, while GHS-BUILT-S measures built-up surface area within each 100 m $\times$ 100 m grid cell. Together, these datasets provide a consistent spatial and temporal basis for examining population distribution, built-up development, and settlement connectivity.

The 50 urban agglomerations were selected using projected 2025 populations from the UN World Urbanization Prospects 2018 Revision~\cite{un2018_2}, ranked in descending order of population. These agglomerations collectively account for approximately 74\% of the population represented by Indian urban agglomerations in that dataset. The corresponding cities and projected populations are listed in Supplementary Table 1. These projections are used only to select the study cities; all subsequent spatial and temporal measurements use GHSL data. India's last completed population census was conducted in 2011, and consistent official spatial boundaries are unavailable for longitudinal analysis of these agglomerations~\cite{populationdata2011}.

For each urban agglomeration, we construct a study region using the administrative districts required to encompass its contiguous settlement footprint. These districts are merged to form an inclusive spatial envelope within which settlement connectivity is analyzed. This approach accommodates metropolitan expansion beyond individual municipal or district boundaries while retaining the surrounding regional settlement fabric. For example, the study region encompassing Delhi comprises 16 districts across four states. Smaller agglomerations may be contained within a single district. The resulting 50 district-based study regions provide the spatial domains for extracting the GHSL rasters across the observation years.

The study-region envelopes are not treated as urban boundaries. Instead, they define the spatial domains within which connected settlement clusters are identified empirically. Consequently, the resulting urban footprints represent subsets of these regions, rather than their complete administrative extent.

\subsection{Spatial connectivity and percolation analysis}

Within each study region and observation year, we identify eligible settlement cells as those containing both positive population and positive built-up surface area. No additional minimum population-density or settlement-size threshold is imposed. The centers of these eligible cells constitute the spatial points used for clustering.

We apply Density-Based Spatial Clustering of Applications with Noise (DBSCAN)~\cite{ester1996dbscan}, following the spatial connectivity principle of the City Clustering Algorithm (CCA)~\cite{rozenfeld2008laws,rozenfeld2011area}. DBSCAN has two parameters: $\mathrm{minPts}$, the minimum number of points within a neighborhood required to form a core point, and $\epsilon$, the maximum distance separating directly connected points.

We use unrestricted DBSCAN ($\mathrm{minPts}=1$), imposing no minimum cluster-size restriction. Under this specification, DBSCAN is equivalent to identifying the connected components of an undirected geometric graph $G(\epsilon)$, whose vertices are eligible settlement cells. Two vertices $i$ and $j$ are connected when the Euclidean distance between their cell centers satisfies

\begin{equation}
\|\mathbf{x}_i-\mathbf{x}_j\|_2 \leq \epsilon,
\end{equation}

where $\mathbf{x}_i$ and $\mathbf{x}_j$ denote the spatial coordinates of the cell centers. Connected components therefore comprise eligible cells linked directly or through sequences of intermediate eligible cells.

For each city-year observation, we systematically vary $\epsilon$ from 100 m to 2,000 m in increments of 50 m. At each sampled length $\epsilon_k$, we identify the largest connected component, containing $S_1(\epsilon_k)$ cells, and calculate its share of the total number of eligible settlement cells, $N$:

\begin{equation}
P(\epsilon_k)=\frac{S_1(\epsilon_k)}{N}.
\end{equation}

As $\epsilon$ increases, additional connections become possible and previously disconnected settlement components may merge. Consequently, $P(\epsilon)$ is non-decreasing. We identify the largest observed connectivity transition as

\begin{equation}
k^*=\underset{k\geq 2}{\arg\max}
\left[
P(\epsilon_k)-P(\epsilon_{k-1})
\right].
\end{equation}

The pre-jump and post-jump connection lengths are defined as $\epsilon_{k^*-1}$ and $\epsilon_{k^*}$, respectively. The corresponding largest connected components provide two alternative morphological delineations of the settlement system immediately before and after its largest observed merger.

Because the analysis concerns finite spatial systems sampled at discrete connection lengths, the largest-jump interval identifies an empirical characteristic merger scale, rather than a universal critical threshold. The 50 m increments determine the resolution at which connectivity changes are observed and do not imply a spatial raster resolution finer than 100 m. Establishing critical phenomena in the statistical-physics sense would require additional scaling analyses beyond the scope of this study.

\subsection{Connectivity robustness and settlement morphology}

To assess whether changes in characteristic connection lengths depend on the largest-jump criterion, we calculate three alternative connectivity thresholds. For a specified proportion $q$ of eligible settlement cells, we define

\begin{equation}
\epsilon_q =
\min\left\{
\epsilon_k :
P(\epsilon_k)\geq q
\right\},
\end{equation}

with $q\in\{0.50,0.75,0.90\}$. These quantities represent the smallest sampled connection lengths at which the largest connected component contains at least 50\%, 75\%, and 90\% of eligible settlement cells, respectively.

We additionally calculate the largest-cluster share at a fixed connection length of 400 m, $P(400\,\mathrm{m})$, and the magnitude of the largest observed merger,

\begin{equation}
\Delta P_{\max}
=
P(\epsilon_{k^*})-
P(\epsilon_{k^*-1}).
\end{equation}

For each measure, annual medians and interquartile ranges are calculated across the 50 cities. Changes between 1975 and 2025 are also evaluated for each city separately, allowing us to report the number of cities exhibiting increases, decreases, or unchanged values, as well as the median of paired city-level changes.

To characterize the spatial organization underlying major connectivity transitions, we map the ten largest connected components immediately before and after the largest observed merger for selected cities. Components are ranked by their number of eligible settlement cells. These maps illustrate differences in the spatial configurations through which previously disconnected settlement components become incorporated into larger connected systems. The resulting morphological interpretations are illustrative rather than mutually exclusive classifications.

\subsection{Population density and comparison with DEGURBA thresholds}

For each pre- and post-jump cluster, total population is calculated by summing GHS-POP values across eligible cells. Urban footprint area is calculated as the number of eligible 100 m $\times$ 100 m cells in the cluster multiplied by 0.01 km$^2$. Cluster population density is consequently defined as

\begin{equation}
D_{\mathrm{cluster}}
=
\frac{P_{\mathrm{cluster}}}
{A_{\mathrm{footprint}}},
\end{equation}

where $P_{\mathrm{cluster}}$ is the total cluster population and $A_{\mathrm{footprint}}$ is its occupied grid-cell area in km$^2$. This measure represents population per unit of connected urban footprint, rather than population per unit of built-up surface.

We compare population densities immediately before and after the largest merger across all 50 cities and eleven observation years, yielding 550 paired city-year observations. We summarize the corresponding distributions using their medians, interquartile ranges, and the proportion of observations exhibiting declining density following the merger.

To examine the relevance of prescribed population-density thresholds, we analyze the distribution of population density across the complete district-based study regions associated with the 50 urban agglomerations. Population counts from GHS-POP are aggregated into non-overlapping 1 km$^2$ grid cells, with density calculated as population per unit grid-cell area. Only cells with positive population are retained. A separate normalized population-density distribution is calculated for each city study region. These 50 distributions are then averaged with equal weight assigned to each region, irrespective of its geographical extent or number of populated grid cells. The resulting distribution is smoothed using a Gaussian filter.

The resulting density distribution is compared with the reference thresholds used by the Degree of Urbanization (DEGURBA) methodology: 300 people/km$^2$ for urban clusters, subject to a minimum total population of 5,000, and 1,500 people/km$^2$ for urban centers, subject to a minimum total population of 50,000~\cite{undesa2025wupmethodology}. These thresholds are used for comparison and are not imposed on the percolation-defined clusters.

%Comment BEGIN UNSDG-REVIEW-03
% Scope of robustness / MAUP (Methods 4.3 and Table 1).
%
% The alternative connectivity summaries support the declining-connection-length finding on the chosen
% geography. They do not test robustness to raster resolution, grid alignment, eligibility thresholds, or
% district-envelope choice. Suggested addition at the end of this subsection:
%
These checks assess sensitivity to the connectivity summary within the selected raster and study regions. They do not establish robustness to the scale or alignment of the underlying grid, the cell-eligibility rule, or the study-region envelope. The resulting delineations and growth estimates remain conditional on these choices \citep{openshaw1977geographical}.
%
% For an additional analysis, compare coarser grids and expanded study envelopes, conserving population and
% built-up totals during aggregation. Report separately whether the direction of connection-length change and
% the expansion/intensification reversal persist. The 100 m cells affect both occupancy and measured
% footprint; changing epsilon affects grouping; district envelopes constrain links and the denominator N.
% These are distinct sources of sensitivity. Do not describe these additional checks as completed unless run.
%
% Reference for the spatial aggregation issue: Openshaw, S. (1981). Le problème de l’agrégation spatiale en
% géographie. L’Espace géographique, 10(1), 15–24. https://doi.org/10.3406/spgeo.1981.3599
%Comment END UNSDG-REVIEW-03

\subsection{Land consumption and population growth}

For each city-year observation, we calculate cluster population, built-up surface area, and urban footprint area separately for the pre- and post-jump connected components. Built-up surface area is defined as the sum of GHS-BUILT-S values within the cluster, converted from m$^2$ to km$^2$. Urban footprint area is defined as the total area of occupied 100 m cells belonging to the connected component.

We estimate land consumption relative to population growth using SDG Indicator 11.3.1, the ratio of land consumption rate to population growth rate (LCRPGR), following the UN-Habitat logarithmic formulation~\cite{UNHabitat2020Metadata1131}. For successive five-year observations, population growth rate and land consumption rate are calculated as

\begin{equation}
\mathrm{PGR}_t =
\frac{\ln(P_t/P_{t-5})}{5},
\qquad
\mathrm{LCR}_t =
\frac{\ln(A_t/A_{t-5})}{5},
\end{equation}

where $P_t$ represents the population of the connected cluster at time $t$, and $A_t$ represents either its built-up surface area or its urban footprint area. The corresponding ratio is

\begin{equation}
\mathrm{LCRPGR}_t
=
\frac{\mathrm{LCR}_t}{\mathrm{PGR}_t}
=
\frac{\ln(A_t/A_{t-5})}
{\ln(P_t/P_{t-5})}.
\end{equation}

The five-year annualization factors cancel in the ratio.

%Comment BEGIN UNSDG-REVIEW-05
% Temporal comparability and changing membership (Methods 4.5; carry qualification into Section 2.4 and
% Discussion).
%
% Holding epsilon constant within each interval is sensible, but independently selecting the largest component
% allows changes in membership and potentially component identity. Suggested addition:
%
% 'Holding the connection length constant does not hold cluster membership constant. Estimated changes can
% therefore reflect both development and the incorporation of settlements already present at the preceding
% date. These comparisons describe the evolution of the delineated settlement system and do not isolate land
% conversion or intensification within a fixed spatial support.'
%
% A fixed-support comparison or component-lineage analysis could quantify the two contributions; please report
% this only if performed. Also state whether study envelopes overlap and, if so, whether any eligible cells or
% connected systems enter more than one city's totals. This matters for the summed 50-city growth percentages
% in the Abstract and Section 2.4. Suggested reporting sentence, once verified: 'For the summed comparisons,
% [state the actual overlap and deduplication rule].'
%
% These are implications of the described measurement procedure, rather than an externally established
% empirical result; no additional citation is needed for the qualification itself.
%Comment END UNSDG-REVIEW-05

To maintain a consistent spatial connection criterion within each growth interval, we use the connection length identified for the current year's pre- or post-jump cluster to delineate the largest connected component independently in both the current and preceding observation years. Thus, temporal comparisons are made at the same connection length rather than between clusters identified using different annual connection lengths. The resulting footprints are nevertheless allowed to change between observations.

For positive population and land-growth rates, LCRPGR values above one indicate that land consumption increases faster than population, values near one indicate approximately proportional growth, and values below one indicate that population increases faster than land consumption. The ratio is undefined when population growth is zero. Nonpositive or undefined LCRPGR values are excluded from logarithmic visualizations, positive-value descriptive summaries, and log-linear scaling regressions. Positive ratios arising from simultaneous declines in population and land area are not interpreted as land-consumptive growth.

We calculate LCRPGR separately for built-up surface and urban footprint, and for pre- and post-jump cluster definitions. City-level trajectories and annual cross-sectional medians are examined over the ten five-year growth intervals between 1980 and 2025. We additionally summarize positive, finite LCRPGR observations across city-period combinations.

\subsection{Rank-size distributions and allometric scaling}

We examine the cross-sectional structure of the 50-city system using rank-size and allometric relationships. For the rank-size analysis, cities are ordered by the population of their largest post-jump connected component in each observation year. We estimate a Zipf-like relationship by ordinary least squares regression of $\ln(\mathrm{rank})$ on $\ln(P)$, reporting the Zipf exponent as the negative of the fitted slope.

To examine the relationship between population and urban land area, we estimate allometric scaling relationships of the form

\begin{equation}
A=aP^\beta,
\end{equation}

where $A$ represents either built-up surface area or urban footprint area, $P$ is cluster population, and $\beta$ is the scaling exponent. We estimate $\beta$ through ordinary least squares regression of $\ln(A)$ on $\ln(P)$, separately for pre- and post-jump clusters and for each observation year. Fitted exponents and coefficients of determination ($R^2$) describe cross-sectional relationships across the 50-city system.

Finally, to investigate whether land consumption relative to population growth varies systematically with city size, we estimate

\begin{equation}
\ln(\mathrm{LCRPGR}) =
\alpha+\beta\ln(P)+\varepsilon,
\end{equation}

separately for built-up surface and urban footprint, and for pre- and post-jump cluster definitions. These regressions include only positive, finite LCRPGR observations, with population measured for the corresponding current-year cluster. We estimate the relationships independently for each five-year growth interval between 1980 and 2025.

The allometric analyses characterize cross-sectional differences between urban systems at a given time, whereas LCRPGR measures longitudinal changes within individual systems. These analyses therefore distinguish the association between city size and urban land area from the rate at which land consumption changes relative to population growth.

\backmatter

\begin{appendices}

\setcounter{figure}{0}
\renewcommand{\thefigure}{\arabic{figure}}

%\bmhead{Supplementary information}

\section*{Supplementary Information}

\subsection*{S1: The 50 largest urban agglomerations in India}

These are arranged in descending order by population size, UN population projections to 2025~\cite{un2018}, extracted from \url{https://population.un.org/wup/DataQuery/}. 

\begin{table}
\centering
    \begin{tabular}{|p{1cm}|p{4cm}|p{4cm}|}
         \hline
          & Cities & 2025 Projection ('000s)  \\
         \hline
         1. & Delhi & 34,666 \\
         \hline
         2. & Mumbai (Bombay) & 22,089 \\
         \hline
         3. & Kolkata (Calcutta) & 15,845 \\
         \hline
         4. & Bangalore & 14,395 \\
         \hline 
         5. & Chennai (Madras) & 12,336 \\
         \hline
         6. & Hyderabad & 11,338 \\
         \hline
         7. & Ahmadabad & 9,062 \\
         \hline
         8. & Surat & 8,582 \\
         \hline
         9. & Pune (Poona) & 7,526 \\
         \hline
         10. & Jaipur & 4,411 \\
         \hline
         11. & Kozhikode (Calicut) & 4,387 \\
         \hline 
         12. & Malappuram & 4,344 \\
         \hline
         13. & Lucknow & 4,133 \\
         \hline
         14. & Thrissur & 3,720 \\
         \hline
         15. & Kochi (Cochin) & 3,605 \\
         \hline
         16. & Indore & 3,483 \\
         \hline
         17. & Kanpur & 3,345 \\
         \hline
         18. & Nagpur & 3,170 \\
         \hline
         19. & Coimbatore & 3,159 \\
         \hline
         20. & Thiruvananthapuram & 3,073 \\
         \hline
         21. & Patna & 2,690 \\
         \hline
         22. & Bhopal & 2,686 \\
         \hline 
         23. & Agra & 2,478 \\
         \hline
         24. & Kannur & 2,465 \\
         \hline
         25. & Visakhapatnam & 2,440 \\
         \hline 
         26. & Vadodara & 2,425 \\
         \hline
         27. & Nashik & 2,351 \\
         \hline
         28. & Vijayawada & 2,351 \\
         \hline
         29. & Kollam & 2,252 \\
         \hline
         30. & Rajkot & 2,150 \\
         \hline
         31. & Ludhiana & 2,029 \\
         \hline
         32. & Raipur & 1,923 \\
         \hline\
         33. & Madurai & 1,912 \\
         \hline
         34. & Meerut & 1,875 \\
         \hline
         35. & Varanasi (Benares) & 1,826 \\
         \hline
         36. & Tiruppur & 1,784 \\
         \hline
        37. & Srinagar & 1,778 \\
        \hline
        38. & Jamshedpur & 1,768 \\
         \hline
        39. & Aurangabad & 1,767 \\
         \hline
        40. & Jodhpur & 1,664 \\
        \hline
        41. & Ranchi & 1,622 \\
        \hline
        42. & Kota & 1,599 \\
        \hline
        43. & Jabalpur & 1,582 \\
        \hline
        44. & Asansol & 1,565 \\
        \hline
        45. & Gwalior & 1,544 \\
        \hline
        46. & Allahabad & 1,524 \\
        \hline
        47. & Amritsar & 1,511 \\
        \hline
        48. & Dhanbad & 1,442 \\
        \hline
        49. & Bareilly & 1,413 \\
        \hline
        50. & Aligarh & 1,380 \\
        \hline
    \end{tabular}
    \caption{50 largest urban agglomerations of India, by UN 2025 population projections}
\label{tab:cities}
\end{table}

\clearpage

\subsection*{S2: Temporal expansion of built-up surface and urban footprint}

\begin{figure}[htbp]
\centering
\includegraphics[width=\textwidth]{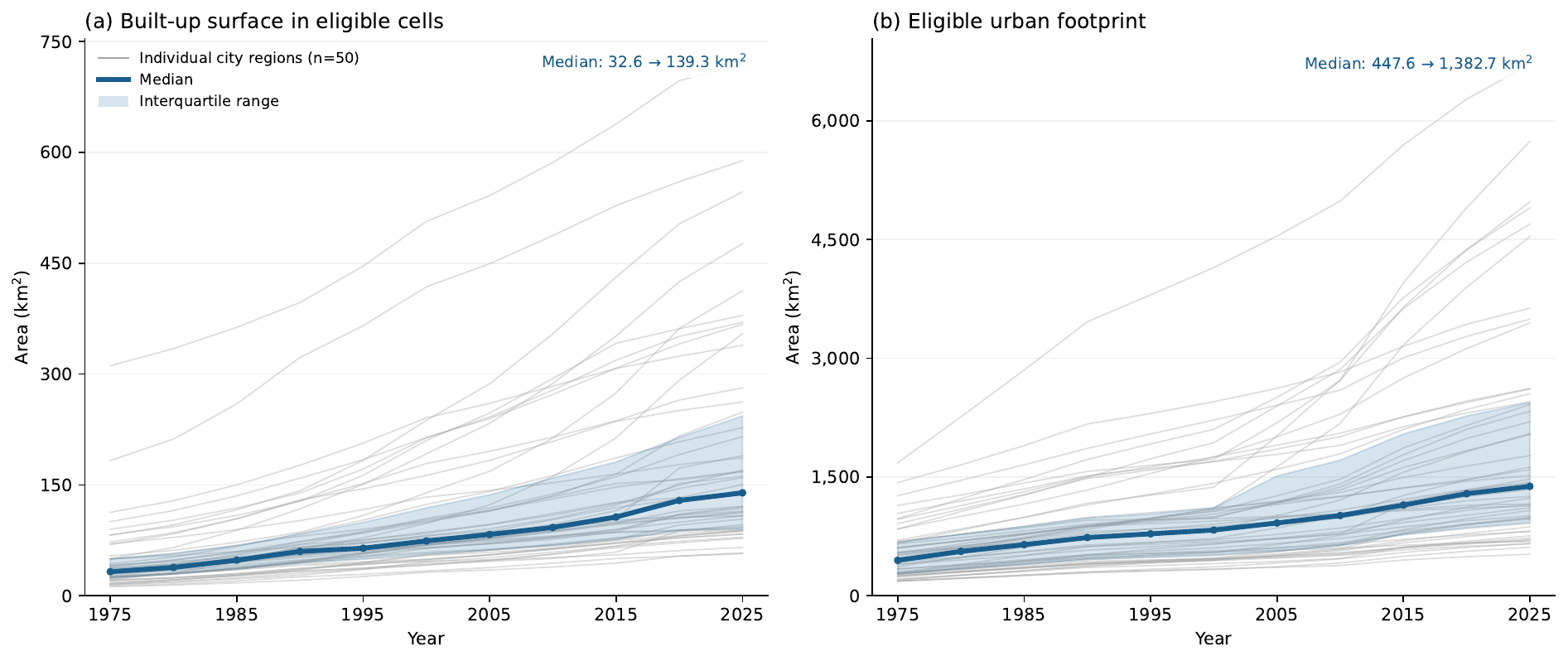}
\caption{\textbf{Temporal expansion of built-up surface and
eligible urban footprint across 50 Indian city study regions,
1975--2025.}
(a) Total built-up surface area within eligible cells and
(b) total eligible urban footprint area across the complete
district-based study regions associated with the 50 urban
agglomerations. Grey lines show individual city study regions,
solid blue lines indicate the cross-sectional median, and
shaded bands represent the interquartile range. Both measures
are expressed in km$^2$ on linear scales. Eligible cells
contain positive population and positive built-up surface.
The study regions are fixed spatial envelopes rather than
percolation-defined urban boundaries.}
\label{fig:supp_full_region_growth}
\end{figure}

We first examine the temporal expansion of built-up surface and
eligible urban footprint across the complete district-based study
regions associated with the 50 urban agglomerations, independently
of the percolation-defined cluster boundaries
(Supplementary Fig.~\ref{fig:supp_full_region_growth}).
Built-up surface is measured as the total GHS-BUILT-S area within
eligible cells, while urban footprint is defined as the total
area of 100 m $\times$ 100 m cells containing both positive
population and positive built-up surface.

Between 1975 and 2025, the median built-up surface area across
the 50 city study regions increased from 32.6 km$^2$ to
139.3 km$^2$, with a median paired city-level increase of
104.8 km$^2$ (interquartile range: 68.8--188.1 km$^2$).
Over the same period, the median number of eligible settlement
cells increased from 44,762 to 138,272, corresponding to an
increase in median eligible urban footprint from 447.6 km$^2$
to 1,382.7 km$^2$. The median paired increase was 88,747
eligible cells, equivalent to 887.5 km$^2$ of additional
eligible urban footprint (interquartile range:
571.7--1,715.4 km$^2$). The median city-level growth
factor in eligible-cell counts was 3.29 (interquartile
range: 2.80--3.92). All 50 city study regions experienced
increases in both built-up surface and eligible urban
footprint between 1975 and 2025. Bangalore recorded the
largest absolute increase in built-up surface
(464.7 km$^2$), while Nashik recorded the largest net
increase in eligible settlement cells (507,450 cells,
equivalent to 5,074.5 km$^2$).

\begin{figure}[htbp]
\centering
\includegraphics[width=\textwidth]{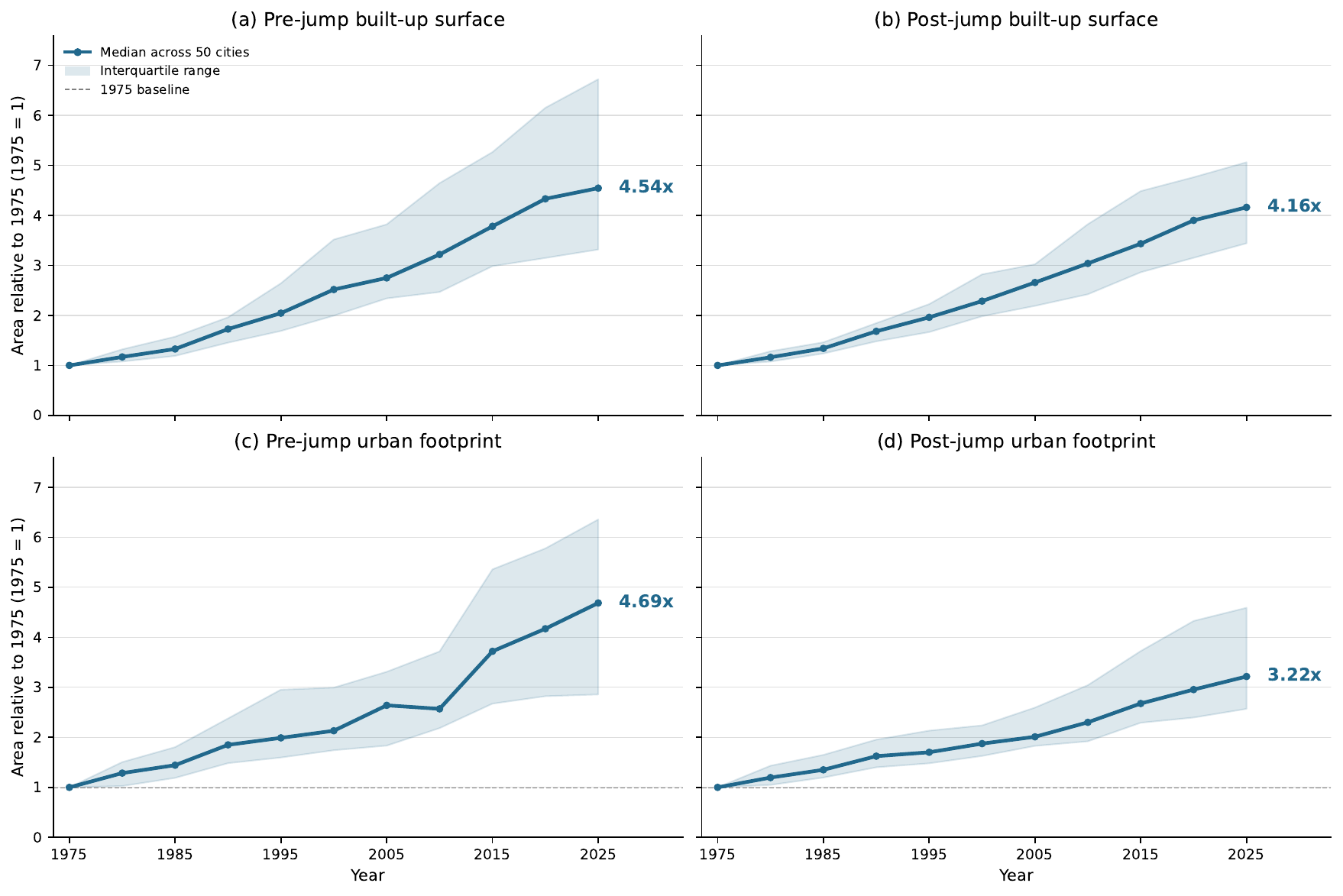}
\caption{\textbf{Relative growth of percolation-defined
connected urban clusters, 1975--2025.}
City-level changes in (a) pre-jump built-up surface,
(b) post-jump built-up surface, (c) pre-jump urban footprint,
and (d) post-jump urban footprint. For each city and
cluster definition, areas are normalized to their
respective 1975 values (1975 = 1). Solid lines indicate
the median across 50 cities, and shaded bands represent
the interquartile range. All panels use linear scales.
Pre- and post-jump components are independently delineated
for each observation year using the characteristic
connection lengths identified by the percolation analysis.}
\label{fig:supp_area_growth}
\end{figure}

We subsequently examine the relative expansion of the
percolation-defined connected components, normalizing each
city's pre- and post-jump cluster areas to their respective
1975 values (Supplementary Fig.~\ref{fig:supp_area_growth}).
By 2025, the median city-level growth factor for built-up
surface was 4.54 (interquartile range: 3.32--6.72) for
pre-jump clusters and 4.16 (3.44--5.06) for post-jump
clusters. The corresponding urban footprint growth factors
were 4.69 (2.86--6.36) and 3.22 (2.57--4.59), respectively.
All 50 cities exhibited positive increases under both
cluster delineations.

These results distinguish the absolute expansion of eligible
settlement fabric within fixed spatial study regions from
the relative growth of percolation-defined connected components.
While the former is independent of the connection length,
changes in the latter may reflect both settlement development
and the incorporation of previously disconnected settlements
through spatial coalescence. Furthermore, increases in eligible
urban footprint measure changes in the extent of populated
built-up cells and should not be interpreted directly as the
conversion of previously undeveloped land.

\subsection*{S3: Temporal changes in population density}

\begin{figure}[htbp]
\centering
\includegraphics[width=\textwidth]{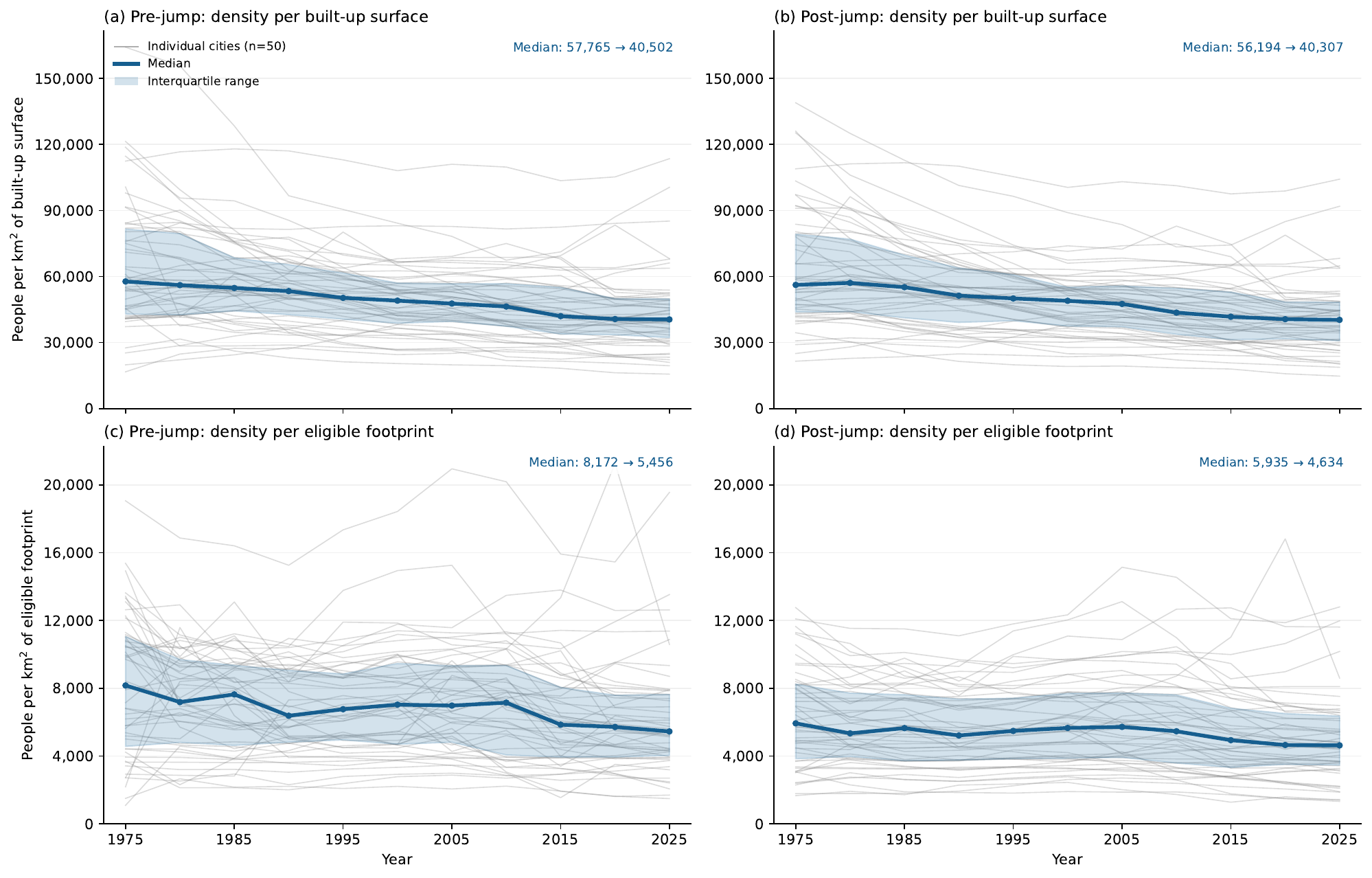}
\caption{\textbf{Temporal evolution of population density
within percolation-defined connected settlement clusters,
1975--2025.}
Population density is calculated as total cluster population
divided by (a) pre-jump built-up surface area,
(b) post-jump built-up surface area,
(c) pre-jump eligible urban footprint area, and
(d) post-jump eligible urban footprint area.
Grey lines represent individual urban agglomerations
($n=50$), solid blue lines indicate the cross-sectional
median, and shaded bands represent the interquartile range.
All densities are expressed in people/km$^2$ on linear
scales. Vertical scales are shared within each row.
Cluster boundaries and characteristic connection lengths
are determined independently for each city and observation
year.}
\label{fig:supp_density_trajectories}
\end{figure}

We examine changes in population density within the pre- and
post-jump connected settlement clusters across the 50 urban
agglomerations between 1975 and 2025
(Supplementary Fig.~\ref{fig:supp_density_trajectories}).
For each city and observation year, population density is
calculated as total cluster population divided by either
the total built-up surface area or the eligible urban
footprint area of the corresponding cluster. These two
measures distinguish population relative to the physical
built-up surface from population relative to the spatial
extent of occupied settlement cells.

Population density per unit of built-up surface declined
substantially over the five decades. For pre-jump clusters,
the cross-sectional median decreased from 57,765 people/km$^2$
in 1975 to 40,502 people/km$^2$ in 2025, while the
corresponding post-jump median declined from 56,194 to
40,307 people/km$^2$
(Supplementary Fig.~\ref{fig:supp_density_trajectories}a,b).
Across paired city observations, 41 of the 50 cities exhibited
declining pre-jump densities, with a median reduction of
25.3\%. For post-jump clusters, densities declined in
43 cities, with a median reduction of 33.5\%.

Population density measured per unit of eligible urban
footprint also declined, although the changes were less
consistent across individual cities. The pre-jump median
decreased from 8,172 people/km$^2$ in 1975 to
5,456 people/km$^2$ in 2025, while the post-jump median
declined from 5,935 to 4,634 people/km$^2$
(Supplementary Fig.~\ref{fig:supp_density_trajectories}c,d).
Declining densities were observed in 34 cities under
pre-jump delineations and 35 under post-jump delineations,
with median paired reductions of 25.0\% and 18.1\%,
respectively.

These findings indicate that the expansion of built-up
surface and eligible urban footprint has generally
outpaced population growth within the percolation-defined
settlement systems. However, the observed density changes
must be interpreted in relation to the evolving spatial
extent of the connected components. In particular, the
incorporation of previously disconnected settlements can
alter aggregate cluster density independently of density
changes within the original urban core. Consequently,
declining cluster-level population density does not
necessarily imply uniform density reductions throughout
individual cities.

\subsection*{S4: Rank-size and allometry analysis reveal sublinear scaling
of built area and urban footprints}

\begin{figure*}[p]
\centering

% Panel (a): Zipf rank-size distribution

\begin{subfigure}[t]{0.70\textwidth}
    \centering
    \includegraphics[
        width=\linewidth,
        height=0.28\textheight,
        keepaspectratio
    ]{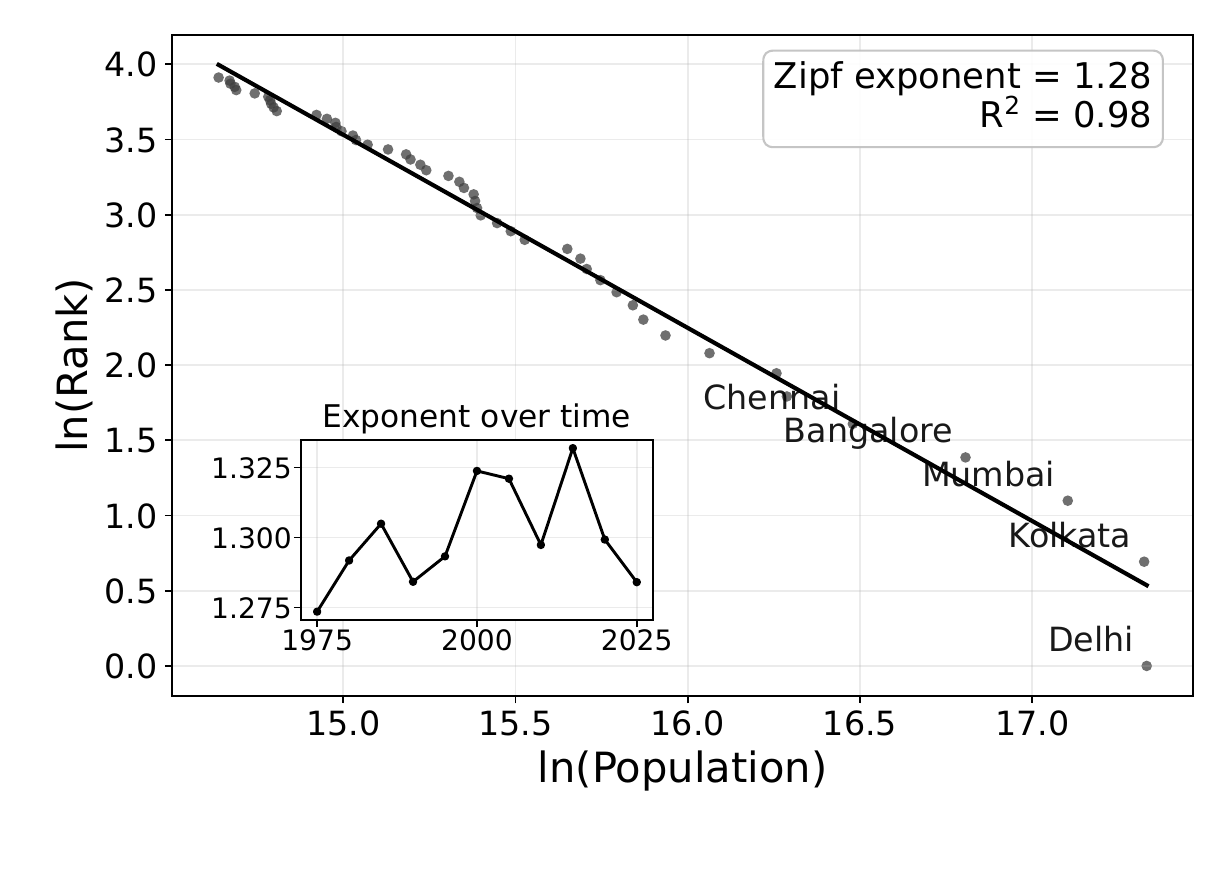}
    \caption{Urban population rank-size distribution}
\end{subfigure}

\vspace{0.5em}

% Panels (b,c): Built-up surface area

\begin{subfigure}[t]{0.48\textwidth}
    \centering
    \includegraphics[
        width=\linewidth,
        height=0.22\textheight,
        keepaspectratio
    ]{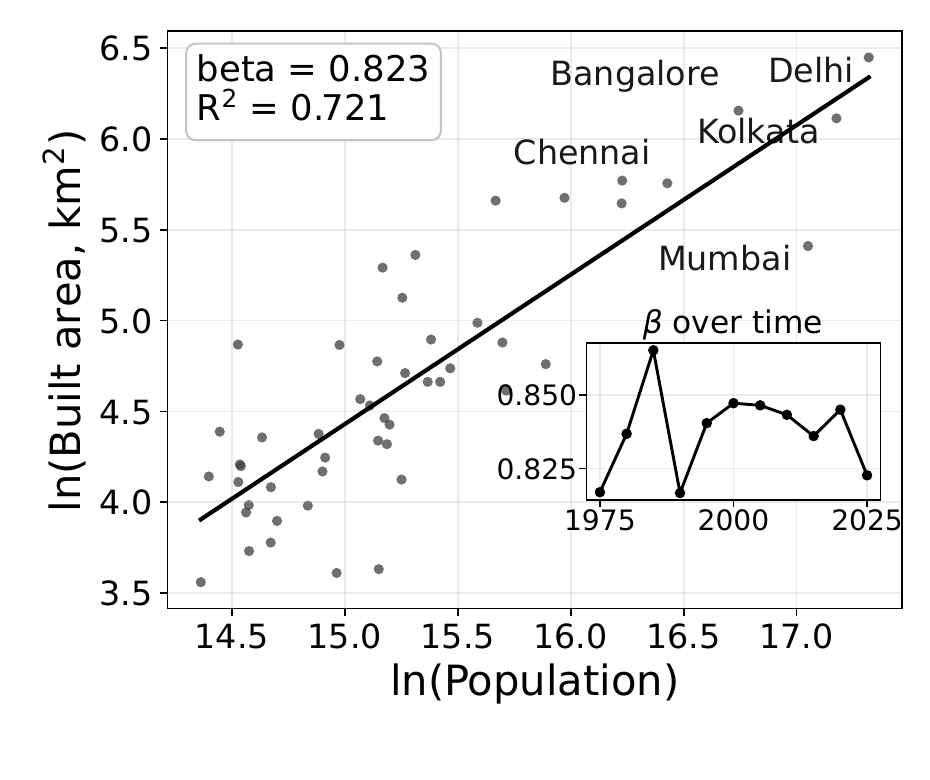}
    \caption{Pre-jump built-up surface area}
\end{subfigure}
\hfill
\begin{subfigure}[t]{0.48\textwidth}
    \centering
    \includegraphics[
        width=\linewidth,
        height=0.22\textheight,
        keepaspectratio
    ]{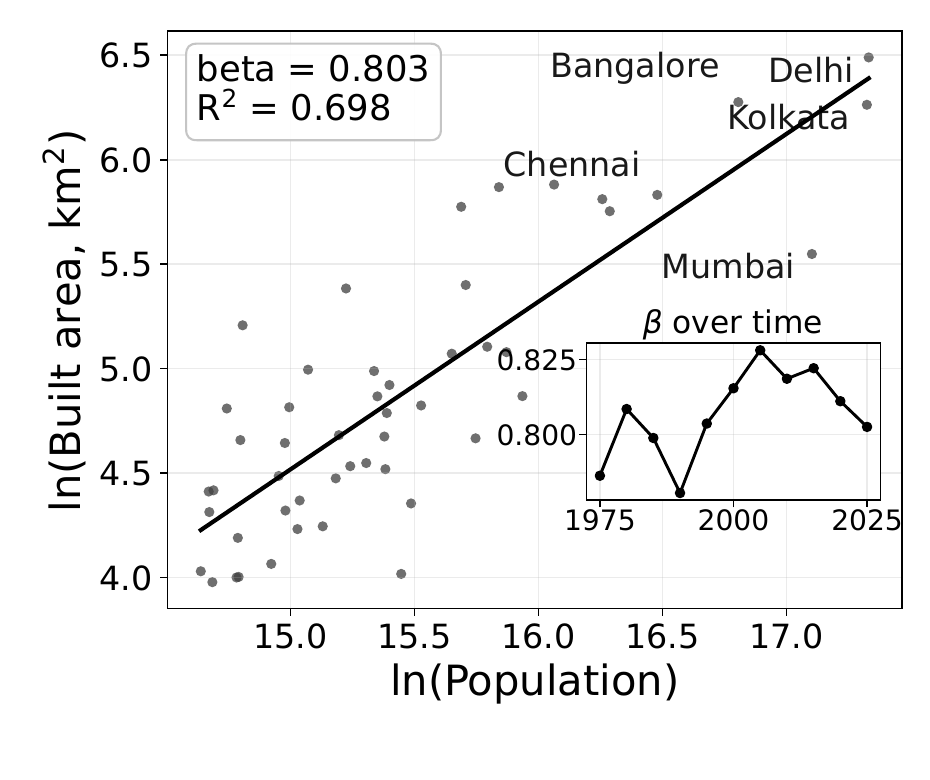}
    \caption{Post-jump built-up surface area}
\end{subfigure}

\vspace{0.5em}

% Panels (d,e): Urban footprint area

\begin{subfigure}[t]{0.48\textwidth}
    \centering
    \includegraphics[
        width=\linewidth,
        height=0.22\textheight,
        keepaspectratio
    ]{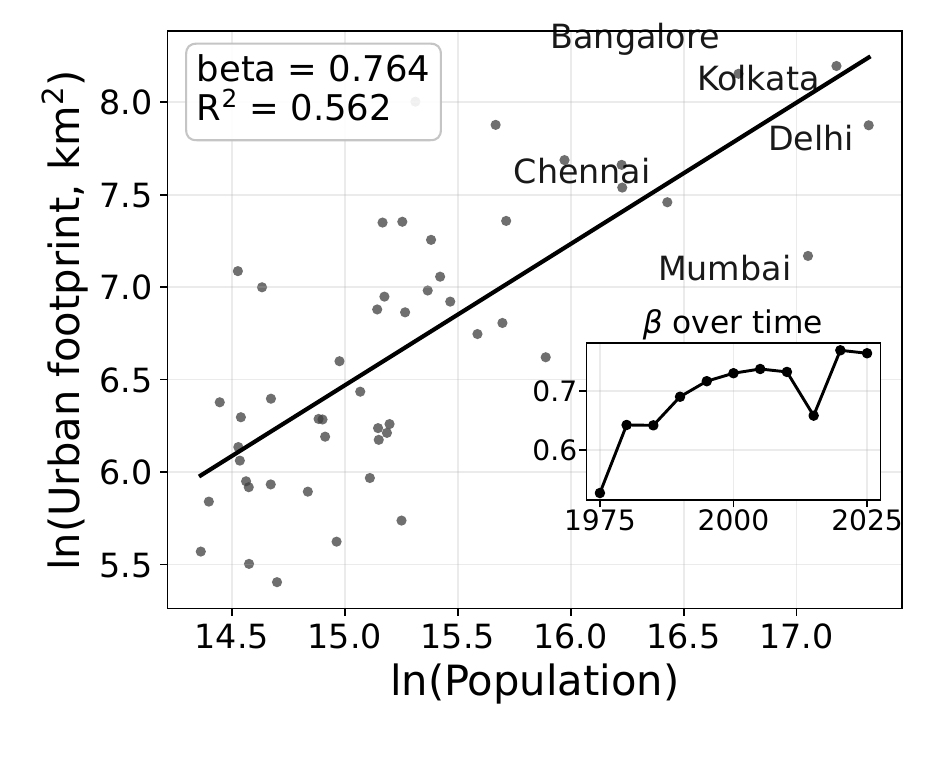}
    \caption{Pre-jump urban footprint area}
\end{subfigure}
\hfill
\begin{subfigure}[t]{0.48\textwidth}
    \centering
    \includegraphics[
        width=\linewidth,
        height=0.22\textheight,
        keepaspectratio
    ]{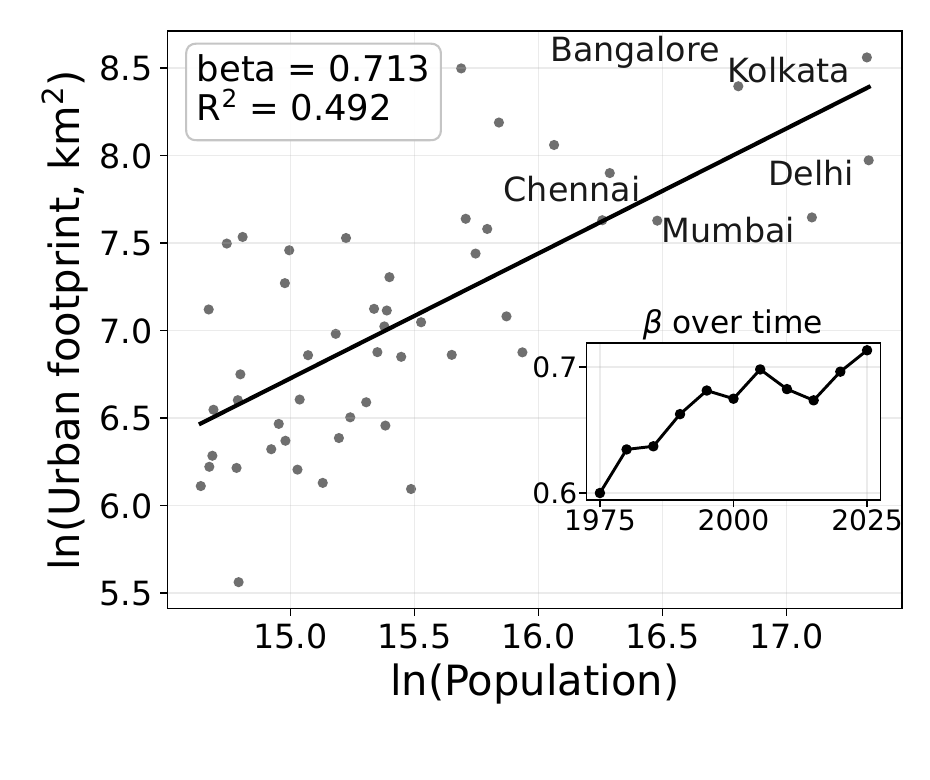}
    \caption{Post-jump urban footprint area}
\end{subfigure}

\caption{
\textbf{Rank-size distributions and sublinear allometric
scaling of percolation-defined urban systems in India.}
(a) Rank-size distribution of the populations of the
largest post-jump connected clusters across 50 Indian
urban agglomerations in 2025. The fitted relationship
between logarithmic population and logarithmic rank
yields a Zipf exponent of 1.28 ($R^2=0.984$).
The inset shows the evolution of the exponent
between 1975 and 2025.
(b,c) Allometric relationships between population
and built-up surface area for pre-jump and post-jump
clusters, respectively.
(d,e) Corresponding relationships between population
and urban footprint area.
Allometric relationships are estimated by ordinary
least squares regression of logarithmic land area
against logarithmic population, with fitted
exponents $\beta$ and coefficients of determination
$R^2$ reported in each panel.
Insets show the temporal evolution of the fitted
allometric exponents between 1975 and 2025.
The consistently sublinear exponents indicate that
larger urban systems occupy less built-up surface
area and connected urban footprint per person
than smaller systems in cross-section.
All measurements use unrestricted DBSCAN
($\mathrm{minPts}=1$).
}
\label{fig:urban_scaling}

\end{figure*}

The percolation-defined settlement system exhibits a strongly ordered urban hierarchy (Supplementary Fig.~\ref{fig:urban_scaling}(a)). In 2025, post-jump cluster populations follow a Zipf-like rank-size relationship, with an exponent of 1.28 ($R^2=0.984$). This is steeper than the classical Zipf benchmark of 1, indicating a less top-heavy hierarchy. The exponent remains relatively stable between 1975 and 2025, ranging from 1.27 to 1.33.

Both built-up surface area and urban footprint scale sublinearly with population across the 50-city system (Supplementary Fig.~\ref{fig:urban_scaling}(b--e)). In 2025, built-up surface scales with exponents of $\beta=0.823$ ($R^2=0.721$) and $\beta=0.803$ ($R^2=0.698$) for pre- and post-jump clusters, respectively. The corresponding urban footprint exponents are $\beta=0.764$ ($R^2=0.562$) and $\beta=0.713$ ($R^2=0.492$). These sublinear relationships indicate that larger urban systems occupy less built-up surface and connected urban footprint per person, on average, in cross-section. Supplementary Fig.~\ref{fig:urban_scaling} also shows the evolution of these exponents from 1975 to 2025.

These cross-sectional patterns do not contradict the longitudinal LCRPGR results. Following the distinction between transversal and longitudinal urban scaling~\cite{MarquisBarthelemy2026}, the allometric exponents describe relationships among cities at a given time, whereas LCRPGR measures land consumption relative to population growth within cities over time. Sublinear cross-sectional scaling can therefore coexist with LCRPGR values above one: larger cities may occupy less land per person than smaller cities at a given time, even as individual cities expand their land area faster than their populations.

%If your article has accompanying supplementary file/s please state so here. 

%Authors reporting data from electrophoretic gels and blots should supply the full unprocessed scans for key as part of their Supplementary information. This may be requested by the editorial team/s if it is missing.

%Please refer to Journal-level guidance for any specific requirements.

%\bmhead{Acknowledgements}

\section*{Declarations}

%Some journals require declarations to be submitted in a standardised format. Please check the Instructions for Authors of the journal to which you are submitting to see if you need to complete this section. If yes, your manuscript must contain the following sections under the heading `Declarations':

\begin{itemize}
\item Funding: The work in this paper was supported by a University of Sydney Deputy Vice Chancellor's UN-SDG Grant, and the University of Sydney and IIT Bombay institutional collaboration. 
\item Conflict of interest/Competing interests: The authors declare no competing interests. 
%\item Ethics approval and consent to participate
%\item Consent for publication
\item Data availability: All data used and generated in this study will be made publicly available in an open-access repository upon publication.
\item Code availability: All code developed for data processing, spatial clustering, statistical analysis, and figure generation will be made publicly available in an open-access repository upon publication.
\item Author contribution: Somwrita Sarkar conceived and designed the study, developed the theoretical and methodological framework, conducted the principal computational and statistical analyses, interpreted the results,
prepared the figures, and wrote the manuscript. Carolina Pantoja
Mardones, Anmol Gaur, Arunim Anand, Rashi Shrivastava, and Anusha
Roy contributed to coding and computational implementation.
Chirag Deb, David Levinson, Arnab Jana, and Eswar Rajasekaran reviewed the manuscript and provided feedback.
\end{itemize}

\end{appendices}


\bibliography{sn-bibliography}% common bib file
%TC:endignore
\end{document}